\documentclass{siamltex}
\usepackage{amsmath,amssymb,mathabx}
\usepackage{graphicx,esint,ulem}
\newcommand{\nc}{\newcommand}

\nc{\fig}[4]
{
  \begin{figure}[ht!]  
    \centering{\scalebox{#1}{\includegraphics*{./figures/#2.eps}}}
    \caption{#4}
    \label{fig:#3}
  \end{figure}
}

\nc{\bXYZ}{{\bf XYZ\ }}

\nc{\nn}{\nonumber} 
\nc{\nit}{\noindent}
\nc{\marginnote}[1] {\marginpar{\tiny #1}}
\nc{\SH}{Schr\"odinger}
\nc{\NLS}{nonlinear Schr\"odinger}
\nc{\ie}{\emph{i.e.\ \mbox{}}}
\nc{\eg}{\emph{e.g.\ \mbox{}}}
\nc{\per}{{l_j}}
\nc{\perx}{q}
\nc{\vol}{{\rm Vol}}
\nc{\Dr}{{\rm D}}
\nc{\notDr}{{{\rm D}^c}}

\nc{\lapx}{\dfrac{\partial^2}{\partial x^2}}
\nc{\lapy}{\dfrac{\partial^2}{\partial y^2}}
\nc{\lapxy}{\dfrac{\partial^2}{\partial xy}}
\nc{\diff}[2]{\frac{d #1}{d #2}}
\nc{\diffn}[3]{\frac{d^{#3} #1}{d {#2}^{#3}}} 
\nc{\pdiff}[2]{\frac{\partial #1}{\partial #2}} 
\nc{\pdiffn}[3]{\frac{\partial^{#3} #1}{\partial{#2}^{#3}}} 

\def\Xint#1{\mathchoice
  {\XXint\displaystyle\textstyle{#1}}%
  {\XXint\textstyle\scriptstyle{#1}}%
  {\XXint\scriptstyle\scriptscriptstyle{#1}}%
  {\XXint\scriptscriptstyle\scriptscriptstyle{#1}}%
  \!\int}
\def\XXint#1#2#3{{\setbox0=\hbox{$#1{#2#3}{\int}$} \vcenter{\hbox{$#2#3$}}\kern-.5\wd0}}
\def\dashint{\Xint-}
\nc{\Av}{\dashint_\cell}
\nc{\avg}[1]{\mbox{$\left \langle\ \!#1\!\!\ \right \rangle$}}
\nc{\intRd}{\int_{{\mathbb R}^d}}

\nc{\abs}[1] {\lvert #1 \rvert} 
\nc{\norm}[2] {{\lVert #1 \rVert}_{#2}} 
\nc{\normH}[1]{\|#1\|_{H^1}^2}
\nc{\normL}[1]{\|#1\|_2^2}
\nc{\nl}[1]{|#1|^{2\sigma}}
\nc{\nlb}[1]{#1^{2\sigma+1}}
\nc{\Linvd}{L_\delta^{-1}}
\nc{\Linvz}{L_0^{-1}}
\nc{\Linvzs}{L_0^{-2}}

\nc{\st}[1]{\stackrel{(\ref{eq:#1})}{=}}
\nc{\stt}[2]{\stackrel{(\ref{eq:#1}),(\ref{eq:#2})}{=}}

\nc{\veps}{\epsilon}
\nc{\Ue}{U_\eps}
\nc{\koe}{\frac{k}{\veps}} 

\nc{\wrat}{w_{\rm ratio}}
\nc{\aij}{A^{ij}}                  
\nc{\minv}{m_*^{-1}} 
\nc{\sqminv}{m_*^{-\frac{1}{2}}}
\nc{\G}{\gamma_{\rm ef\,\!f}}
\nc{\zetap}{{\zeta_*}}
\nc{\zetas}{{\zeta_{1*}}}
\nc{\Pe}{P_{\rm edge}}
\nc{\slope}{\dfrac{d\cP[u(\cdot;\mu)] }{d\mu}}
\nc{\moot}{\mu_2}
\nc{\mum}{\mu_{\rm min}}

\def\R{\mathbb{R}}
\def\C{\mathbb{C}}
\def\Z{\mathbb{Z}}

\nc{\brill}{{\mathcal{B}}}
\nc{\cell}{{\Omega}}
\nc{\bj}{{\bf j}}
\nc{\bk}{{\bf k}}
\nc{\bl}{{\bf l}}
\nc{\bm}{{\bf m}}
\nc{\bn}{{\bf n}}
\nc{\bkappa}{{\mathbf \kappa}}
\nc{\bK}{{\bf K}}
\nc{\bKp}{{\bf K'}}
\nc{\bp}{{\bf p}}
\nc{\bpi}{{\mathbf{\pi}}}
\nc{\bq}{{\bf q}}
\nc{\br}{{\bf r}}
\nc{\bv}{{\bf v}}
\nc{\bx}{{\bf x}}
\nc{\bX}{{\bf X}}
\nc{\bY}{{\bf Y}}
\nc{\by}{{\bf y}}
\nc{\bz}{{\bf z}}
\nc{\bnu}{{\bf \nu}}
\nc{\bxc}{{\bf x}_c}
\nc{\bxi}{{\bold\xi}}

\nc{\cP}{{\cal P}}
\nc{\cPedge}{{\cal P}_{edge}}
\nc{\cPc}{{\cal P}_{cr}} 
\nc{\cDD}{\Lambda} 
\nc{\cF}{{\cal F}}
\nc{\cG}{{\cal G}}
\nc{\cO}{{\cal O}}  
\nc{\cQ}{{\cal Q}}  
\nc{\cR}{{\cal R}}
\nc{\cI}{{\cal I}}
\nc{\cK}{{\cal K}}
\nc{\cL}{{\cal L}} 
\nc{\cM}{{\cal M}}
\nc{\cN}{{\cal N}} 
\nc{\cE}{{\cal E}}
\nc{\cH}{{\cal H}} 
\nc{\cX}{{\cal X}}
\nc{\cZ}{{\cal Z}} 
\nc{\cT}{{\cal T}} 
\nc{\order}{{\cal O}}
\nc{\tT}{{\tilde T}} 
\nc{\tlambda}{{\tilde\lambda}} 

\newtheorem{remark}{Remark}[section]

\newcommand{\x}{\mathbf{x}}

\newcommand{\eps}{\varepsilon}

\newcommand{\D}{\partial}
\newtheorem{thm}{Theorem}[section]

\nc{\pulse}{%
  \begin{picture}(60,50)(0,-20)
  \qbezier(0,0)(5,0)(10,15)
  \qbezier(10,15)(15,30)(20,30)
  \qbezier(20,30)(22,30)(25,18)
  \qbezier(25,18)(27,10)(30,10)
  \qbezier(30,10)(34,10)(38,15)
  \qbezier(38,15)(42,20)(45,20)
  \qbezier(45,20)(49,20)(53,10)
  \qbezier(53,10)(57,0)(60,0)
  \multiput(0,0)(0,-2){10}{\line(0,-1){1}}
  \multiput(60,0)(0,-2){10}{\line(0,-1){1}}
  \put(30,-20){\vector(-1,0){30}}
  \put(30,-20){\vector(1,0){30}}
  \put(0,-15){\makebox(60,10){{\tiny $T$}}}
  \end{picture}}

\begin{document}

\title{Wave packets in Honeycomb Structures and Two-Dimensional Dirac Equations }

\author{Charles L. Fefferman\footnotemark[1] and  Michael I. Weinstein\footnotemark[2]}

\maketitle

\renewcommand{\thefootnote}{\fnsymbol{footnote}}
\footnotetext[1]{Department of Mathematics, Princeton University; cf@math.princeton.edu}
\footnotetext[2]{Department of Applied Physics and Applied
  Mathematics, Columbia University; miw2103@columbia.edu}

\maketitle

\begin{abstract}
In a recent article \cite{FW:12}, the authors proved that the non-relativistic Schr\"odinger operator with a  generic honeycomb lattice potential has conical (Dirac) points in its dispersion surfaces. These conical points occur for quasi-momenta, which are located at the vertices of the Brillouin zone, a regular hexagon. In this paper, we  study the time-evolution of wave-packets, which are spectrally concentrated near such conical points. We  prove that the large, but finite, time dynamics
is  governed by the two-dimensional  Dirac equations.  
 \end{abstract}
%

\begin{keywords}
Dirac equation, Honeycomb Lattice Potential, Graphene,  Floquet-Bloch theory, Dispersion Relation
\end{keywords}
%
%
\pagestyle{myheadings}
\thispagestyle{plain}
\markboth{Effective Dirac Dynamics }{Charles L. Fefferman and Michael I. Weinstein}

\section{Introduction and Outline}
\label{sec:introduction-outline}
\medskip

There is great interest within the fundamental and applied physics communities in the properties of waves in periodic structures with honeycomb lattice symmetry. The (Floquet-Bloch) dispersion relation of such structures is known to have conical singularities which occur at the intersections of certain bands at high-symmetry quasi-momenta.
 These conical singularities, also called  {\it Dirac points} or {\it diabolical points}, are central to the remarkable electronic properties of graphene \cite{RMP-Graphene:09,goerbig:11} and wave-propagation properties in dielectrics (linear and nonlinear) with honeycomb structure dielectric parameters \cite{Haldane:08,Segev-etal:07,Segev-etal:08,Rechtsman-etal:12} . Conical points have long been known to arise in the dispersion relation of plane waves of the homogeneous and {\it anisotropic} Maxwell equations \cite{Berry-Jeffrey:07}.
 
 In  \cite{FW:12}  it was proved that for generic honeycomb lattice potentials, $V(\bx)$, that the non-relativistic time-independent Schr\"odinger equation:
  \begin{equation}
\mu \phi= \left(-\Delta +V(\bx)\right)\phi
 \label{t-inde-schroedinger}
 \end{equation}
  has conical singularities in its dispersion surfaces. These occur at quasi-momenta located at the vertices of the Brillouin zone, $\brill$, a regular hexagon.  In this paper we prove that the dynamics of solutions of the time-dependent Schr\"odinger equation:
  \begin{equation}
 i\D_t\psi = \left(-\Delta +V(\bx)\right)\psi\ ,
 \label{td-schroedinger}
 \end{equation}
   for initial data which are spectrally concentrated near the vertices of $\brill$,
   are for very large, but finite, times effectively governed by a two-dimensional system of relativistic Dirac equations. We next explain our main result, Theorem \ref{effective-Dirac}.\medskip
   
    It is natural to decompose solutions of \eqref{td-schroedinger} in terms of its Floquet-Bloch states: $\Phi_b(\bx;\bk)e^{-i\mu_b(\bk)t}$,
 where $b\ge1$, $\bk\in\brill$ and $\mu_b(\bk),\ b\ge1$ are the  eigenvalues of the pseudo-periodic eigenvalue problem with quasi-momentum, $\bk$; see \eqref{phi-eqn}-\eqref{pseudo-per}.  At a conical singularity (Dirac point) of a honeycomb structure we have two dispersion surfaces, graphs of consecutive maps $\bk\mapsto \mu_{b_1}(\bk)\equiv\mu_-(\bk)$ and 
  $\bk\mapsto \mu_{b_1+1}(\bk)\equiv\mu_+(\bk)$ intersecting conically at each vertex, $\bK_\star$ of the Brillouin zone, $\brill$:\  $\mu_\star\equiv
\mu_+(\bK_\star)=\mu_-(\bK_\star)$. The $b_1^{th}$ and $(b_1+1)^{st}$ spectral bands intersect at the  energy $\mu_\star$ and this energy is attained by $\mu_-(\bk)$ and $\mu_+(\bk)$ at each of the vertices of $\brill$. The corresponding two-dimensional  quasi-periodic eigenspace associated with the quasi-momenta $\bK_\star$,
${\rm Nullspace}\left(-\Delta+V-\mu_\star\right)$, is spanned by the pair: $\Phi_1(\bx;\bK_\star)$ and $\Phi_2(\bx;\bK_\star)$, which satisfy the relation: $\Phi_2(\bx)=\overline{\Phi_1(-\bx)}$; see the notion of  Dirac point, Definition \ref{Diracpt-def}. 

   Theorem \ref{effective-Dirac} asserts the following for a generic honeycomb lattice potential, $V(\bx)$:\   Consider initial conditions
   of the form:
    \begin{align}
\psi(\bx,0)\ &=\  \sum_{j=1}^2\ \delta\ \alpha_{j0}(\delta\bx)\Phi_j(\bx)\ ,
\label{wavepacket-data}\end{align}
with fixed, smooth, rapidly decreasing $\alpha_{j0}(\bX),\ j=1,2$ and $\delta$ small. 
We call this a {\it wave-packet} spectrally localized at 
 $\bK_\star\in\mathcal{B}$. For such initial conditions the solution evolves, approximately, as a slowly modulated superposition of Floquet-Bloch states:
 \begin{align}
\psi(\bx,t)\ &\approx e^{-i\mu_\star t}\ \sum_{j=1}^2\ \delta\ \alpha_j(\delta\bx,\delta t)\ \Phi_j(\bx),
\label{ansatz}\end{align}
where the modulating amplitudes, $\alpha_j(\bX,T)$, satisfy the effective Dirac system
\begin{align}
 \D_T\alpha_1(\bX,T)\ &=\ -\overline{\lambda_\sharp}\ \left(\D_{X_1}+i\D_{X_2}\right)\ \alpha_2(\bX,T)\label{Dirac-1}\\ 
\D_T\alpha_2(\bX,T)\ &=\ -\lambda_\sharp\ \left(\D_{X_1}-i\D_{X_2}\right)\ \alpha_1(\bX,T)\  ,
\label{Dirac-2}
\end{align}  
where $0\ne\lambda_\sharp\in\mathbb{C}$.
\ \  In Theorem \ref{effective-Dirac} we establish the validity of \eqref{ansatz} where $\alpha_1,\alpha_2$ satisfy \eqref{Dirac-1}-\eqref{Dirac-2},  on  time scales of order $\mathcal{O}(\delta^{-2+\eps})$, for any $\eps>0$.
\medskip

To prove Theorem \ref{effective-Dirac}, we seek a solution of the initial value problem with wave-packet initial condition \eqref{wavepacket-data} with leading order term given by the right hand side of \eqref{ansatz} plus a correction term, $\eta^\delta(x,t)$, which is represented via the DuHamel formula; see \eqref{eta-eqn}-\eqref{eta-integ-eqn}. The Dirac equations \eqref{Dirac-1}-\eqref{Dirac-2} arise as a non-resonance condition, which ensures that $\eta^\delta(\bx,t)$ is small on a time interval: $0\le t\le\mathcal{O}(\delta^{-2+\eps})$, for any $\eps>0$. Estimation of $\eta^\delta$  requires a careful decomposition of the propagator, $e^{-i(-\Delta+V)t}$ and analysis of its action on functions with quasi-momentum components supported near $\bK_\star$, a vertex of $\brill$, and those with quasi-momentum components supported away from $\bK_\star$. The resonant terms which are removed by imposing equations \eqref{Dirac-1}-\eqref{Dirac-2}, arise from quasi-momenta near $\bK_\star$. A detailed expansion of the normalized Floquet-Bloch modes for such quasi-momenta is required. Such modes are discontinuous at  $\bK_\star$. Components corresponding to quasi-momenta away from $\bK_\star$ are controlled, via Poisson summation and integration by parts with respect to time, by making use of rapid phase oscillations in time.
\medskip

Formal derivations of Dirac-type dynamics for honeycomb lattice structures are discussed in the physics   \cite{RMP-Graphene:09} and  applied mathematics \cite{ANZ:09,Ablowitz-Zhu:11} literature. A rigorous discussion of the tight-binding limit is presented in \cite{Ablowitz-Curtis-Zhu:12}. 
Conical singularities have long been known to occur in Maxwell equations with constant anisotropic dielectric tensor;
  see, for example, \cite{Indik-Newell:06},  \cite{Berry-Jeffrey:07} and references cited therein. 
 \bigskip
 
To put our results in context, we discuss the effective dynamics of two other classes of initial conditions:

\begin{enumerate}
\item {\bf Ballistic propagation}\  \cite{Allaire-05}:\  Take data given by a  wave-packet which is localized at a frequency, 
$\widetilde{\mu}\ = \mu_{\widetilde b}(\widetilde\bK)$, 
where $\mu_{\widetilde{b}}(\bk)$ is regular in a neighborhood of $\widetilde{\bK}$, $\widetilde\mu$ is a simple eigenvalue of $H(\widetilde\bK)$ with corresponding Floquet-Bloch eigenstate $\Phi_{\widetilde b}(\bx;\widetilde\bK)$ and $\nabla_\bk\mu_{\widetilde b}(\widetilde\bK)\ne{\bf 0}$:
\[ \psi_0(\bx,0)\ =\ \delta\ \alpha_0(\delta\bx)\ \Phi_{\widetilde b}(\bx;\widetilde\bK)\ .\]
Then, the large time approximate evolution is given by:
\begin{align}
&\psi(\bx,t)\ \approx\ e^{-i\widetilde\mu t}\ \delta\ \alpha(\delta\bx,\delta t)\ \Phi_{\widetilde b}(\bx;\widetilde\bK) \nn\\
& \D_T\alpha(\bX,T)\ +\ \nabla_\bk\mu_{\widetilde b}(\widetilde\bK)\cdot\nabla_\bX\ \alpha(\bX,T)=0,\ \ \bX=\delta\bx,\ T=\delta t\nn
\end{align}
Thus, 
\begin{equation}
 \psi(\bx,t)\ \approx\ e^{-i\widetilde\mu t}\
 \delta\ \alpha_0\Big(\ \delta\cdot \left(\ \bx-\nabla_\bk\mu_{\widetilde b}(\widetilde\bK)\ t\ \right) \ \Big)\ \Phi_{\widetilde b}(\bx;\widetilde\bK) 
\label{ballistic}\end{equation} 
for times, $t$, of order $\delta^{-2}$.
\item 
{\bf  Effective mass (homogenized) Schr\"odinger evolution \cite{Allaire-05}:}\ Let $\widetilde\bK$ be such that $\mu_{\widetilde b}(\widetilde\bK)$ occurs at a spectral band (gap) edge. Take wave-packet data which is spectrally localized near the frequency $\mu_{\widetilde b}(\widetilde\bK)$:
\[ \psi_0(\bx,0)\ =\ \delta\ \alpha_0(\delta\bx)\ \Phi_{\widetilde b}(\bx;\widetilde\bK),\ \ 0<\delta\ll1.\]
Since $\mu_{\widetilde b}(\widetilde\bK)$ is at a band edge,  we have $\nabla_\bk\mu_{\overline b}(\widetilde\bK)={\bf 0}$. Furthermore,
 assume the Hessian matrix $D^2_\bk\mu_{\tilde b}(\widetilde\bK)$ is non-degenerate. Then, the large time approximate evolution is given by:
\begin{align}
&\psi(\bx,t)\ \approx\ e^{-i\widetilde\mu t}\ \delta\ \alpha(\delta\bx,\delta^2 t)\ \Phi_{\widetilde b}(\bx;\widetilde\bK) 
\nn\end{align}
where $\alpha(\bX,\tau)$ is governed by the constant coefficient Schr\"odinger equation:
\begin{align}
& i\D_\tau\alpha(\bX,\tau)\ = -\nabla_\bX\cdot A_{\rm eff}\nabla_\bX \alpha(\bX,\tau),\ \ \ \bX=\delta\bx,\ \ T=\delta^2t
\label{effective-mass-schrod}\\
& A_{\rm eff} =\ \frac{1}{2}\ D^2_\bk\mu_{\widetilde b}(\widetilde\bK)\nn
\end{align}
for times, $t$,  of the order $\delta^{-2}$. $A_{\rm eff}$ is referred to as the inverse of the effective mass tensor.
\end{enumerate}
\bigskip

\subsection{Outline of the paper}\label{outline}\  In Section 
\ref{sec:spectr-theory-peri} we review basic Floquet-Bloch theory for general periodic potentials and introduce the class of honeycomb lattice potentials. In Section \ref{Dirac-points} we discuss the main results of the authors' recent paper \cite{FW:12} as well as some direct consequences required in the current work. In Section \ref{2D-Dirac-eqn} we discuss properties of solutions to the two-dimensional Dirac system \eqref{Dirac-1}-\eqref{Dirac-2}. In Section \ref{sec:wave-packet} we state our main result, Theorem \ref{effective-Dirac} on the large, but finite, time Dirac effective dynamics for appropriate wave-packet initial data for the time-dependent Schr\"odinger equation with a generic honeycomb lattice potential. The proof of Theorem \ref{effective-Dirac} is contained in Sections \ref{pf-effectiveD} and \ref{proof-of-fast-avg}. Appendix \ref{EigLip} gives an elementary proof of the Lipschitz continuity of eigenvalues of self-adjoint operators. We thank B. Simon for a sketch of a shorter proof using standard perturbation theory; see Chapter XII of \cite{RS4}.

In a forthcoming article, we  present  an analytic perturbation theory of deformed honeycomb lattice Hamiltonians, for perturbations which commute with inversion composed with complex conjugation. Conical (Dirac) points persist for small  perturbations of this type, although the conical singularities typically perturb away from the vertices of $\brill$. These results extend those of \cite{FW:12} and, in particular, include the case of a uniformly strained honeycomb structure.  We also consider the analogous question of the dynamics of solutions for wave-packet initial data, spectrally concentrated at a Dirac point of the deformed honeycomb structure. In this case, the methods of the present article apply to establish the large, but finite, time dynamics as being given by {\it tilted-} Dirac equations.  The latter can be mapped to the standard 2D Dirac equations by a Galilean change of variables. 

 \bigskip

\noindent {\bf Acknowledgements:}\ 
CLF was supported by US-NSF Grant DMS-09-01040.
 MIW was supported in part by US-NSF Grant DMS-10-08855.  The authors wish to thank  M. Ablowitz, A.C. Newell and G. Uhlmann for stimulating discussions. 
 \bigskip

\subsection{Notation} 
\label{sec:notation}
%
%
\begin{enumerate}
\item $z\in\mathbb{C}\ \implies\ \overline{z}$ denotes the complex conjugate of $z$.
\item $A$, a $d\times d$ matrix $\implies$ $A^t$ is its transpose and $A^*$ is its conjugate-transpose.
\item $\bK^\bm=\bK^{m_1,m_2}=\bK+\bm\bk=\bK+m_1\bk_1+m_2\bk_2$.\\  $\bK, \bk_1$ and $\bk_2$ are defined in Section \ref{sec:honeycomb}.
\item $\brill$ denotes the standard Brillouin zone of Figure \ref{fig:honeyAB-2}. $\brill_h$ denotes an equivalent choice, introduced for convenience in the proofs, centered at $\bK$.
\item $\nabla_\bk=e^{-i\bk\cdot\bx}\nabla_\bx e^{i\bk\cdot\bx}=\nabla_\bx+i\bk$,\ $\Delta_\bk=\nabla_\bk\cdot\nabla_\bk$.
\item ${\bf x}, {\bf y}\in\C^n,\ \ \left\langle {\bf x}, {\bf y}\right\rangle=\overline{\bf x}\cdot{\bf y}$,\ 
$\bx\cdot \by=x_1y_1+\dots +x_ny_n$.
\item For $\bq=(q_1,q_2)\in\mathbb{Z}^2$, $\bq\bk=q_1\bk_1+q_2\bk_2$\ .
\item $\langle f,g\rangle=\int \overline{f}g$
\item $x\lesssim y$ if and only if there exists $C>0$ such that $x\le Cy$. 
\item We write $f=\mathcal{O}_X(\rho)$ if there exists a constant, $C$, such that $\|f\|_X\le C\rho$.
\end{enumerate}

\section{Periodic Potentials and Honeycomb Lattice Potentials}
\label{sec:spectr-theory-peri}

We begin with a review of Floquet-Bloch theory of periodic potentials \cite{Eastham:73,Wilcox:78,Kuchment-01,RS4}.

\subsection{Floquet-Bloch Theory} 

Let $\{\bv_1,\bv_2\}$ be a linearly independent set in $\mathbb{R}^2$. Consider the lattice 
\begin{equation}
\Lambda=\{m_1\bv_1+m_2\bv_2: m_1,m_2\in\Z\ \}=\Z\bv_1\oplus\Z\bv_2\ .
\label{Lambda-def}
\end{equation}
The fundamental period cell is denoted 
\begin{equation}
\Omega=\{\ \theta_1\bv_1+\theta_2\bv_2: 0\le\theta_j\le1,\ j=1,2\ \}\ .
\label{Omega-def}
\end{equation}
Denote by $L^2_{per,\Lambda}\ =\ L^2(\R^2/\Lambda)$, the space of  $L^2_{\rm loc}$ functions which are periodic with the respect to the lattice $\Lambda$, or equivalently functions in $L^2$ on the torus
 $\mathbb{R}^2/\Lambda=\mathbb{T}^2$:
\begin{equation}
f\in L^2_{per,\Lambda} \ \textrm{if and only if}\  \ f(\bx+\bv)=f(\bx), \ {\rm for}\ \bx\in\R^2 ,\ \ \bv\in\Lambda\ .
\nn\end{equation}
More generally, we consider functions satisfying a pseudo-periodic boundary condition:
\begin{equation}
f\in L^2_{\bk,\Lambda}  \ \textrm{if and only if}\  \ f(\bx+\bv)=f(\bx)e^{i\bk\cdot\bv}, \ {\rm for}\ \bx\in\R^2 ,\ \ \bv\in\Lambda .
\end{equation}
We shall suppress the dependence on the period-lattice, $\Lambda$, and write $L^2_\bk$,  if the choice of lattice is clear from context. For $f$ and $g$ in $L^2_{\bk,\Lambda}$, $\overline{f}g$ is locally integrable and $\Lambda$- periodic and we define their inner product by:
\begin{equation}
\left\langle f,g\right\rangle\ =\ \int_\Omega\ \overline{f(\bx)}\ g(\bx)\ d\bx\ .
\label{inner-product}
\end{equation}
In a standard way, one can introduce the Sobolev spaces $H^s_{\bk,\Lambda}$.

The dual lattice, $\Lambda^*$, is defined to be
\begin{equation}
\Lambda^*\ =\ \{m_1\bk_1+m_2\bk_2 : m_1, m_2\in\mathbb{Z}\}=\Z\bk_1\oplus\Z\bk_2 \ ,
\label{eq:dual-lattice}
\end{equation}
where $\bk_1$ and $\bk_2$ are dual lattice vectors, satisfying the relations:
\begin{equation}
\bk_i\cdot \bv_j = 2\pi \delta_{ij}\ .
\nn\end{equation}

If $f\in L^2_{per,\Lambda}$ then  $f$ can be expanded in a Fourier series with Fourier coefficients
 $\hat{f} = \{ f_\bm \}_{\bm\in\mathbb{Z}^2}$:
\begin{align}
f(\bx)\ &=\ \sum_{\bm\in\Z^2} f_\bm\ e^{i\bm\bk\cdot\bx}\ =\ \sum_{(m_1,m_2)\in\Z^2} f_{m_1,m_2}\ e^{i(m_1\bk_1+m_2\bk_2)\cdot \bx}\ \ \ \ ,\\
f_\bm\ & \equiv\ \ \frac{1}{|\cell|}\ \int_{\cell}\ e^{-i\bm\bk
\cdot \by}\ f(\by)\ d\by\ =\ \frac{1}{|\cell|}\ \int_{\cell}\ e^{-i(m_1\bk_1+m_2\bk_2)
\cdot \by}\ f(\by)\ d\by.
\label{Fourier-coeff}\end{align}

 \bigskip
Let $V(\bx)$ denote a real-valued potential which is periodic
 with respect to $\Lambda$, {\it i.e.}
\begin{equation} V(\bx+\bv)=V(\bx), \ {\rm for}\ \bx\in\R^2 ,\ \ \bv\in\Lambda\ .
\nn\end{equation}
Throughout this paper we shall also assume that
\begin{equation}
V\in C^\infty(\mathbb{R}^2/\Lambda)\ .
\label{Vassumptions}
\end{equation}
We expect that this smoothness assumption can be relaxed considerably without much extra work.

For each $\bk\in\R^2$ we consider the  {\it Floquet-Bloch eigenvalue problem}
\begin{align}
H_V\  \Phi(\bx;\bk) &= \mu(\bk)\ \Phi(\bx;\bk),\ \ \bx\in\R^2,\label{phi-eqn}\\
 \Phi(\x+\bv;\bk) &= e^{i\bk\cdot \bv}\ \Phi(\bx;\bk),\ \ \bv\in \Lambda,\   \label{pseudo-per}
 \end{align}
 where
 \begin{equation}
 H_V\ \equiv\  -\Delta + V(\bx)\ .
 \label{HV-def}
 \end{equation}
 An $L^2_\bk$- solution of \eqref{phi-eqn}-\eqref{pseudo-per} is called a {\it Floquet-Bloch} state. A function which satisfies the boundary condition \eqref{pseudo-per} is said to be $\bk-$ pseudo-periodic.
 \medskip

 Since the eigenvalue problems \eqref{phi-eqn}-\eqref{pseudo-per} are invariant under the change $\bk\mapsto \bk+\tilde{\bk}$, where $\tilde{\bk}\in\Lambda^*$, the dual period lattice, the eigenvalues and eigenfunctions of  \eqref{phi-eqn}-\eqref{pseudo-per} can be regarded as  $\Lambda^*-$ periodic functions of $\bk$, or functions on $\mathbb{R}^2/\Lambda^*$.
  Therefore, it suffices to restrict our attention to $\bk$ varying over any primitive cell.
  It is standard to work with the first Brillouin zone, $\mathcal{B}$, the closure of the set of  points $\bk\in\mathbb{R}^2$, which are closer to the origin than to any other lattice point.

 An alternative formulation is obtained as follows. For every $\bk\in\mathcal{B}$ we express the Floquet-Bloch mode, $\Phi(x;\bk)$,  in the form
 \begin{equation}
 \Phi(\bx;\bk)=e^{i\bk\cdot\bx}p(\bx;\bk)\ . \label{phi-p}
 \end{equation}
Then $p(\bx;\bk)$  satisfies the periodic elliptic boundary
value problem: 
\begin{align}
H_V(\bk) p(\bx;\bk)\ &=\   \mu(\bk)\ p(\bx;\bk),\ \  \bx\in\R^2,\label{p-eqn}\\
  p(\bx+\bv;\bk) &= p(\bx;\bk),\ \ \bv\in \Lambda \label{p-periodic}, 
\end{align}
where
\begin{equation}
H_V(\bk)\equiv -\left(\nabla+ i\bk\right)^2 + V(\bx)\ .
\label{HVdef}
\end{equation}
The eigenvalue problem \eqref{phi-eqn}-\eqref{pseudo-per}, or equivalently \eqref{p-eqn}-\eqref{p-periodic},  has a discrete spectrum:
\begin{equation}
\mu_1(\bk)\ \le\ \mu_2(\bk)\ \le\ \mu_3(\bk)\ \le\ \dots
\label{eig-ordering}\end{equation}
with eigenpairs
$
p_b(\bx;\bk),\ \mu_b(\bk):\ b=1,2,3,\dots .
$
The set $\{p_b(\bx;\bk)\}_{b\ge1}$ can be taken to be a complete orthonormal set in 
$L^2_{\rm per}(\R^2/\Lambda)$.  
\medskip

The functions $\mu_b(\bk)$ are called band dispersion functions. Some general results on their regularity appear in \cite{Wilcox:78, avron-simon:78}.

Since $V$ is assumed to be smooth, elliptic regularity theory  implies  for each $b\ge1$ and $\bk\in\brill$, that
\begin{align}
\textrm{ $\bx\mapsto p_b(\bx;\bk)$ is $C^\infty(\R^2/\Lambda)$.}
\nn\end{align}
Furthermore, there exists a constant $C_{b,\beta,V}$, depending only on $b, \beta$ and $V$,  such that
\begin{equation}
\max_{\bk\in\brill}\ \|\D_\bx^\beta p_b(\cdot;\bk)\|_{L^\infty(\Omega)}\ <\ C_{b,\beta,V} .
\label{Dpb}\end{equation}
\medskip

We shall also require the regularity of the mapping $\bk\mapsto\mu_b(\bk)$. \smallskip

\begin{proposition}\label{ktomuk}
 The eigenvalue maps $\bk\mapsto\mu_b(\bk),\ b\ge1$, are Lipschitz continuous.
\end{proposition}

Proposition \ref{ktomuk} (see also Proposition \ref{lip-continuity-fb-eigs}) is a consequence of the general result on Lipschitz continuity of eigenvalues of second order elliptic operators (Theorem \ref{lip-continuity}), stated and proved in Appendix \ref{EigLip}.\medskip

\begin{remark}\label{nasty-evectors}
Although the eigenvalue maps, $\bk\mapsto\mu_b(\bk)$, are Lipschitz functions
the Floquet-Bloch mode maps, $\bk\mapsto p_b(\bx;\bk)$, are in general not even continuous \cite{Wilcox:78}. Indeed, we shall see this behavior explicitly in a neighborhood of  degenerate eigenvalues; see Theorem \ref{p-near-kD}.
\end{remark}

As $\bk$ varies over $\mathcal{B}$, $\mu_b(\bk)$ sweeps out a closed real interval. The spectrum of $-\Delta + V(\bx)$ in $L^2(\R^2)$ is the union of these closed
intervals:
\begin{equation}
  \label{L2-spectrum}
  \textrm{spec}(H_V) = \bigcup_{ \bk \in \mathcal{B}} \textrm{spec}\left(H_V(\bk)\right) \ \ \ .
\end{equation}
Moreover, the set 
\begin{equation}
\bigcup_{b\ge1}\bigcup_{\bk\in\mathcal{B}}\{\Phi_b(\bx;\bk)\},
\qquad \Phi_b(\bx;\bk)\equiv e^{i\bk\cdot\bx}p_b(\bx;\bk),\label{completeset}
\end{equation}
  suitably normalized, is complete in $L^2(\R^2)$:
%
\begin{align}
 f(\bx) &=\sum_{b\ge1} \int_{\mathcal{B}}\ \left\langle \Phi_b(\cdot;\bk),f(\cdot)\right\rangle_{L^2(\mathbb{R}^2)}\ \Phi_b(\bx;\bk)\ d\bk\ =\ \sum_{b\ge1} \int_{\mathcal{B}}\ \tilde{f}_b(\bk)\ \Phi_b(\bx;\bk)\ d\bk\ ,
\label{f-expanded}
\end{align}
where the sum converges in the $L^2$ norm.
\medskip

Moreover we have, with respect to the Floquet-Bloch basis, the Plancherel Theorem:
\begin{align}
\|f\|_{L^2(\R^2)}^2\ &=\ \sum_{b\ge1} \int_{\mathcal{B}}\ |\tilde{f}_b(\bk)|^2\ d\bk
\label{plancherel}
\end{align}
\medskip

\begin{remark}\label{recenterBZ}
The $\Lambda_h^*-$ periodicity of the Floquet-Bloch modes implies that we can express \eqref{f-expanded} equivalently in terms of a $d\bk-$ integral over any fundamental period cell. A convenient choice, to be used below, is one where 
 the integal over $\brill$ is replaced by an integral over 
 \begin{equation}
 \brill_h\ \equiv\ \bK\ +\ \brill\ .
 \label{brill-def}
 \end{equation}
  That $\bK$ is an interior point to this fundamental domain, rather than a vertex, will simplify certain computations below.
\end{remark}
\medskip

Thus it is natural to introduce Sobelev spaces, defined in terms of the Floquet-Bloch coefficients as follows:
\begin{align}
\|f\|_{H^s(\R^2)}^2\ &\approx\ \|(I+|H|^2)^\frac{s}{4}f\|_{L^2(\R^2)}^2\ =\ 
\sum_{b\ge1} \int_{\mathcal{B}}\ (1+|\mu_b(\bk)|^2)^{s\over2}\ |\tilde{f}_b(\bk)|^2\ d\bk\nn\\
&\approx \sum_{b\ge1} (1+|b|^2)^{s\over2}\ \int_{\mathcal{B}}\ |\tilde{f}_b(\bk)|^2\ d\bk\ . 
\label{Hs}\end{align}
The latter approximation is a consequence of:
\begin{equation}
 |\mu_b(\bk)|\sim |b|,\ \ b\gg1\ .
 \label{Weyl}
 \end{equation}
 The Weyl law \eqref{Weyl} holds uniformly in $\brill$.

Note the simple consequence of  \eqref{Hs}, to be used in Section \ref{proof-of-fast-avg}:
\begin{equation}
 \tilde{f}_b\equiv0,\ \textrm{ for $b$ outside a fixed finite set}\ \implies\ 
 \|f\|_{H^s(\R^2)}\lesssim\|f\|_{L^2(\R^2)},\ \ s\ge0.
 \label{cor-Hs}\end{equation}

\subsection{The period lattice, $\Lambda_h$ , and its dual, $\Lambda_h^*$}
\label{sec:honeycomb}
{\ \ \ \ }\bigskip

\begin{figure}[ht!]
\centering \includegraphics[width=4.75in]{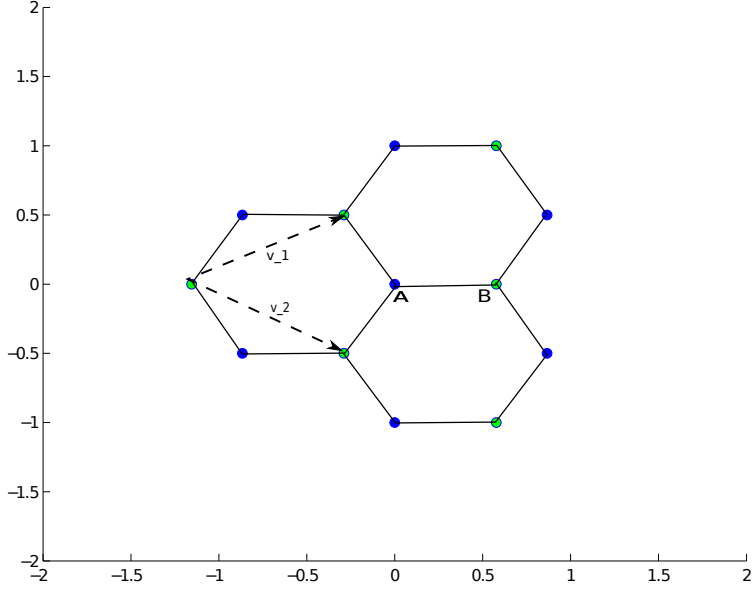}
  \caption{ Part of the honeycomb structure, ${\bf H}$.
  ${\bf H}$ is the union of two sub-lattices  $\Lambda_{\bf A}={\bf A}+\Lambda_h$ (blue) 
   and $\Lambda_{\bf B}={\bf B}+\Lambda_h$ (green). The lattice vectors 
    $\{\bv_1,\bv_2\}$ generate $\Lambda_h$. 
   } .  \label{fig:honeyAB-1}
\end{figure}

Consider $\Lambda_h=\Z{\bf v}_1 \oplus \Z{\bf v}_2$, the lattice generated by the basis vectors:
\begin{align}
 {\bf v}_1 &=\ a\left( \begin{array}{c} \frac{\sqrt{3}}{2} \\ {}\\  \frac{1}{2}\end{array} \right),\ \ 
{\bf v}_2 =\ a\left(\begin{array}{c} \frac{\sqrt{3}}{2} \\ {}\\ -\frac{1}{2} \end{array}\right),\ \ a>0.\label{v12-def}
\end{align}
Note: $\Lambda_h$ (``$h$'' for honeycomb) is a triangular lattice, that arises naturally in connection with honeycomb structures; see Figure \ref{fig:honeyAB-1}.

 \begin{figure}[ht!]
\centering 
\includegraphics[width=4.75in]{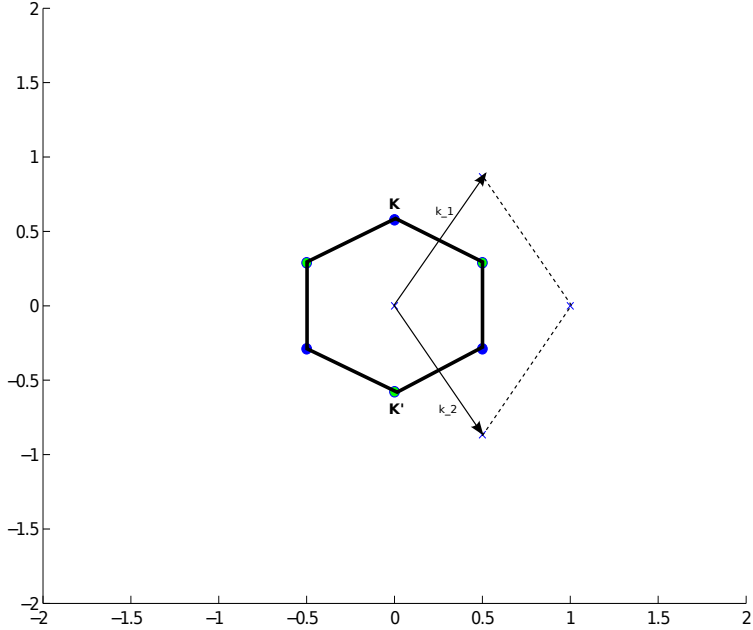}
  \caption{  Brillouin zone, $\brill$, and dual basis $\{\bk_1,\bk_2\}$. $\bK$ and $\bK'$ are labeled. Other vertices of $\brill_h$ obtained via application of $R$, rotation by $2\pi/3$; see equation \eqref{Bvertices}.} .  \label{fig:honeyAB-2}
\end{figure}

The dual lattice $\Lambda_h^* =\ \Z {\bf k}_1\oplus \Z{\bf k}_2$ is spanned by the dual basis vectors:
\begin{align}
&  {\bf k}_1=\ q\left(\begin{array}{c} \frac{1}{2}\\ {}\\ \frac{\sqrt{3}}{2}\end{array}\right),\ \ \ {\bf k}_2 = q\ \left(\begin{array}{c}\frac{1}{2}\\ {}\\ -\frac{\sqrt{3}}{2}\end{array}\right),\ \  \ q\equiv \frac{4\pi}{a\sqrt{3}}\ ,\label{q-def}
\end{align}
where
\begin{align}
&{\bf k}_{\ell}\cdot {\bf v}_{{\ell'}}=2\pi\delta_{\ell\ell'}\ , \label{orthog-kv}\\
&|\bv_1|=|\bv_2|=a,\ \ \bv_1\cdot\bv_2=\frac{a^2}{2}\ , \label{bv12}\\
&|\bk_1|=|\bk_2|=q,\ \ \bk_1\cdot\bk_2=-\frac{1}{2}q^2\ .\label{bk12}
\end{align}

The Brillouin zone, $\brill$, is a hexagon in $\R^2$; see Figure \ref{fig:honeyAB-2}. Denote by $\bK$ and $\bKp$ the  vertices of  $\brill$ given by:
\begin{equation}
\bK\equiv\frac{1}{3}\left(\bk_1-\bk_2\right),\ \ \bKp\equiv-\bK=\frac{1}{3}\left(\bk_2-\bk_1\right)\ .
\label{KKprime}
\end{equation}
All six  vertices of $\brill$ can be generated by application of the rotation matrix, $R$,
 which rotates a vector in $\mathbb{R}^2$ clockwise by $2\pi/3$. $R$ is given by
\begin{equation}
R\ =\ \left(\begin{array}{cc} -\frac{1}{2} & \frac{\sqrt{3}}{2}\\
                                                 {} & {}\\
                                           -\frac{\sqrt{3}}{2} & -\frac{1}{2}
                                           \end{array}\right)
  \label{Rdef}\end{equation}
  and the vertices of $\brill$ fall into two groups, generated by action of $R$ on $\bK$ and $\bK'$:
\begin{align}
&\bK\ {\rm type-points:}\ \bK,\ R\bK=\bK+\bk_2,\ R^2\bK=\bK-\bk_1\nn\\
&\bKp\ {\rm type-points:}\ \bKp,\ R\bKp=\bK'-\bk_2,\ R^2\bKp=\bKp+\bk_1\ .
\label{Bvertices}
\end{align}
\begin{remark}[Symmetry Reduction]\label{symmetry-reduction}
Let $(\ \Phi(\bx;\bk), \mu(\bk)\ )$ denote a Floquet-Bloch eigenpair for the eigenvalue problem \eqref{phi-eqn}-\eqref{pseudo-per} with quasi-momentum $\bk$. Since $V$ is real, 
$(\ \tilde{\Phi}(\bx;\bk)\equiv\overline{\Phi(\bx;\bk)}, \mu(\bk)\ )$ is a Floquet-Bloch eigenpair for the eigenvalue problem with quasi-momentum $-\bk$. Recall the relations \eqref{Bvertices}  and the $\Lambda_h^*$- periodicity of: $\bk\mapsto\mu(\bk)$ and $\bk\mapsto \Phi(\bx;\bk)$.    It follows that  the local character of the dispersion surfaces in a neighborhood of any vertex of $\brill$ is determined by its character about any other vertex of $\brill$.
\end{remark}
\bigskip

We reminder the reader that below, it will be convenient to work with $\brill_h=\bK+\brill$ as our Brillouin zone, explained in Remark \ref{recenterBZ}.

\subsection{Honeycomb lattice potentials}\label{sec:HLP}
{\ \ \ }

For any function $f$, defined on $\mathbb{R}^2$,  introduce
\begin{equation}
\mathcal{R}[f](\bx)=f(R^*\bx),\label{calRdef}
\end{equation}
where $R$ is the $2\times2$ rotation matrix displayed in \eqref{Rdef}.
\medskip

\begin{definition}[Honeycomb lattice potentials]\label{honeyV}
{ }

Let $V$ be  real-valued and  $V\in C^\infty(\R^2)$.
$V$ is  a  \underline{ honeycomb lattice potential} 
 if there exists $\bx_0\in\mathbb{R}^2$ such that $\tilde{V}(\bx)=V(\bx-\bx_0)$ has the following properties:
\begin{enumerate}
\item $\tilde{V}$ is $\Lambda_h-$ periodic, {\it i.e.}  $\tilde{V}(\bx+\bv)=\tilde{V}(\bx)$ for all $\bx\in\mathbb{R}^2$ and $\bv\in\Lambda_h$.  
\item $\tilde{V}$ is even or inversion-symmetric, {\it i.e.} $\tilde{V}(-\bx)=\tilde{V}(\bx)$.
\item  $\tilde{V}$  is $\mathcal{R}$- invariant, {\it i.e.}
 \begin{equation}
 \mathcal{R}[\tilde{V}](\bx)\ \equiv\ \tilde{V}(R^*\bx)\ =\ \tilde{V}(\bx),
 \nn\end{equation}
  where, $R^*$ is the counter-clockwise rotation matrix by $2\pi/3$, {\it i.e.} $R^*=R^{-1}$, where $R$
 is given by \eqref{Rdef}. 
 
 \end{enumerate}\smallskip
 
 Thus, a honeycomb lattice potential is smooth, $\Lambda_h$- periodic and, with respect to some origin of coordinates, both inversion symmetric and $\mathcal{R}$- invariant.
 \end{definition}\medskip
 
 \begin{remark}\label{takex0eq0}
 As the spectral properties are independent of translation of the potential we shall assume in the proofs, without any loss of generality,  that $\bx_0=0$.
 \end{remark}\medskip
 
 \begin{remark}\label{Vreal-even}
 A consequence of a honeycomb lattice potential being real-valued and even is that if 
 $\left(\Phi(\bx;\bk),\mu\right)$ is an eigenpair with quasimomentum $\bk$ of the Floquet-Bloch eigenvalue problem, then $\left(\overline{\Phi(-\bx;\bk)},\mu\right)$ is also an eigenpair with quasimomentum $\bk$.
 \end{remark} 
 \medskip
 
 A key property of honeycomb lattice potentials, $V_h$, used in our spectral analysis 
  of $-\Delta+V_h$ \cite{FW:12}, is that if $\bK_\star$ denotes any vertex of $\brill_h$, then we have the commutation relation:
  \begin{equation}
  \left[\mathcal{R},H_{V_h}(\bK_\star)\right]\ =\ 0\ .
  \label{commutation}
  \end{equation}
  It is therefore natural to the split $L^2_{\bK_\star}$, the space of $\bK_\star$- pseudo-periodic functions, into the direct sum:
  \begin{equation}
   L^2_{\bK_\star}\ =\ L^2_{\bK_\star,1}\oplus L^2_{\bK_\star,\tau}\oplus L^2_{\bK_\star,\overline\tau},\label{L2-directsum}
   \end{equation}
   where $L^2_{\bK_\star,\sigma}$ are invariant eigen-subspaces of $\mathcal{R}$, {\it i.e.}
  for $\sigma=1,\tau,\overline{\tau}$, where $\tau=\exp(2\pi i/3)$, and
   \begin{equation}
   L^2_{\bK_\star,\sigma}\ =\ \Big\{g\in  L^2_{\bK_\star}: \mathcal{R}g=\sigma g\Big\}\ .
\label{L2Ksigma}   \end{equation}

\section{ Dirac points}\label{Dirac-points}

{\qquad}\medskip

We begin with a precise definition of a Dirac point. 
\begin{definition}\label{Diracpt-def}
Let $V(\bx)$ be a smooth,  real-valued, even (inversion symmetric) and  periodic potential on $\R^2$.
Denote by $\brill$ the Brillouin zone given in Remark \ref{recenterBZ}.
We call $\bK\in\brill$ a \underline{Dirac point} if the following holds:
There exist an integer $b_1\ge1$, a real number $\mu_\star$, and strictly positive  numbers, $\lambda$ and $\delta$, such that:
\begin{enumerate}
\item $\mu_\star$ is a degenerate eigenvalue of $H$ with $\bK-$ pseudo-periodic boundary conditions.
\item $\textrm{dim Nullspace}\Big(H-\mu_\star I\Big)\ =\ 2$
\item $\textrm{Nullspace}\Big(H-\mu_\star I\Big)\ =\ 
{\rm span}\Big\{ \Phi_1(\bx) , \Phi_2(\bx)\Big\}$, where $\Phi_1\in L^2_{\bK,\tau}$ and $\Phi_2(\bx)=\overline{\Phi_1(-\bx)}\in L^2_{\bK,\bar\tau}$. 
\item There exist Lipschitz functions $\mu_\pm(\bk)$, \[ \mu_{b_1}(\bk)=\mu_-(\bk)\ \ \  
\mu_{b_1+1}(\bk)=\mu_+(\bk),\ \ \mu_\pm(\bK)=\mu_\star\]
 and $E_\pm(\bk)$, defined for $|\bk-\bK|<\delta$, and 
  $\bk-$ pseudo-periodic eigenfunctions of $H$: $\Phi_\pm(\bx;\bk)$, with corresponding eigenvalues $\mu_\pm(\bk)$ such that
  \begin{align}
\mu_+(\bk)-\mu_\star\ &=\ +\ \lambda\ 
\left| \bk-\bK \right|\ 
\left(\ 1\ +\ E_+(\bk)\ \right)\ \ {\rm and}\nn\\
\mu_-(\bk)-\mu_\star\ &=\ -\ \lambda\ 
\left| \bk-\bK \right|\ 
\left(\ 1\ +\ E_-(\bk)\ \right),\label{cones}
\end{align}
where $|E_\pm(\bk)|\le C|\bk-\bK|$ for some $C>0$. 
  \end{enumerate}
\end{definition}
\medskip

\begin{remark}\label{lambda-lambdasharp}
 In \cite{FW:12} we prove the following\medskip
 
\begin{proposition}\label{lambda-is=|lambdasharp|}
Suppose conditions $1. , 2.$ and $3.$ of Definition \ref{Diracpt-def} hold and denote by 
$\{c(\bm)\}_{\bm\in\mathcal S}$ the sequence of $L^2_{\bK,\tau}$ Fourier-coefficients of $\Phi_1(\bx)$. 
Define the sum
\begin{equation}
\lambda_\sharp\ \equiv\   \sum_{\bm\in\mathcal{S}} c(\bm)^2\ \left(\begin{array}{c}1\\ i\end{array}\right)\cdot \bK_\star^\bm\ \ ,
\label{lambda-sharp}
\end{equation}
with $\mathcal{S}\subset\Z^2$; see \cite{FW:12}.
If $\lambda_\sharp\ne0$, then  
$4.$ of Definition \ref{Diracpt-def} holds (see \eqref{cones} )  with $\lambda=|\lambda_\sharp|$.\\
\end{proposition}
\end{remark}

%
\medskip

We next recall the statement of Theorem 5.1 of \cite{FW:12} concerning the existence of Dirac points for the Schr\"odinger operator with a generic  honeycomb lattice potentials.

\begin{thm}\label{main-thm} 
Let $V_h(\bx)$ honeycomb lattice potential. Assume further that the Fourier coefficient of $V_h$, $V_{1,1}$, is non-vanishing, {\it i.e.}
\begin{equation}
V_{1,1}\ \equiv\ \int_\Omega e^{-i(k_1+k_2)\cdot\by}\ V_h(\by)\ d\by\ \ne0\ .
\label{V11eq0}
\end{equation}
Consider the one-parameter family of honeycomb Schr\"odinger operators defined by:
\begin{equation}
H^{(\epsilon)}\ \equiv\ -\Delta + \epsilon\ V_h(\bx)\ .
\label{Heps-def}
\end{equation}
There exists a countable and 
closed set $\tilde{\mathcal{C}}\subset\mathbb{R}$ such that for    all $\epsilon\notin\tilde{\mathcal{C}}$, the vertices, $\bK_\star$, of $\brill_h$ are Dirac points in the sense of Definition \ref{Diracpt-def}.\medskip

More specifically,  the following holds for $\epsilon\notin\tilde{\mathcal{C}}$:
 There exists $b_1\ge1$ such that $\mu_\star\equiv\mu^\epsilon_{b_1}(\bK_\star)=\mu^\epsilon_{b_1+1}(\bK_\star)$ is a $\bK_\star-$ pseudo-periodic eigenvalue of multiplicity two
  where
\begin{enumerate}
\item   
\subitem $\mu^\epsilon_\star$  is an $L^2_{\bK,\tau}$ - eigenvalue of $H^{(\epsilon)}$ of multiplicity one, and corresponding eigenfunction, 
$\Phi^\epsilon_1(\bx)$.
\subitem $\mu^\epsilon_\star$ is an $L^2_{\bK,\bar\tau}$ - eigenvalue of $H^{(\epsilon)}$ of multiplicity one, with corresponding eigenfunction, $\Phi^\epsilon_2(\bx)=\overline{\Phi^\epsilon_1(-\bx)}$.
\subitem $\mu^\epsilon_\star$ is \underline{not} an $L^2_{\bK,1}$- eigenvalue of $H^{(\epsilon)}$.
\item There exist $\delta_\epsilon>0,\ C_\epsilon>0$
  and Floquet-Bloch eigenpairs: $(\Phi_+^\epsilon(\bx;\bk), \mu_+^\epsilon(\bk))$  
 and $(\Phi_-^\epsilon(\bx;\bk), \mu_-^\epsilon(\bk))$, and  Lipschitz continuous functions, $E_\pm(\bk)$,   defined for  $|\bk-\bK_\star|<\delta_\epsilon$,  such that 
\begin{align}
\mu^\epsilon_+(\bk)-\mu^\epsilon(\bK_\star)\ &=\ +\ |\lambda^\epsilon_\sharp|\ 
\left| \bk-\bK_\star \right|\ 
\left(\ 1\ +\ E^\epsilon_+(\bk)\ \right)\ \ {\rm and}\nn\\
\mu^\epsilon_+(\bk)-\mu^\epsilon(\bK_\star)\ &=\ -\ |\lambda^\epsilon_\sharp|\ 
\left| \bk-\bK_\star \right|\ 
\left(\ 1\ +\ E^\epsilon_-(\bk)\ \right),\nn
\end{align}
where 
\begin{equation}
\lambda_\sharp^\epsilon\ \equiv\   \sum_{\bm\in\mathcal{S}} c(\bm,\mu^\epsilon,\epsilon)^2\ \left(\begin{array}{c}1\\ i\end{array}\right)\cdot \bK_\star^\bm\ \ne\ 0
\label{lambda-sharp2}
\end{equation}
 is given in terms of $\{c(\bm,\mu^\epsilon,\epsilon)\}$, the $L^2_{\bK,\tau}$- Fourier coefficients of $\Phi^\epsilon(\bx;\bK_\star)$.

Furthermore, $|E_\pm^\epsilon(\bk)| \le C_\epsilon |\bk-\bK_\star|$. 
Thus, in a neighborhood of the point $(\bk,\mu)=(\bK_\star,\mu_\star^\epsilon)\in \R^3 $, the dispersion surface is closely approximated by a  circular {\it cone}.
\item 
There exists $\epsilon^0>0$, such that for all $\epsilon\in(-\epsilon^0,\epsilon^0)\setminus\{0\}$\\
  (i)\  $\epsilon V_{1,1}>0\ \implies$ conical intersection of $1^{st}$ and $2^{nd}$ dispersion surfaces \\
 (ii)\ $\epsilon V_{1,1}<0\ \implies$ conical intersection of $2^{nd}$ and $3^{rd}$ dispersion surfaces\ .
 \end{enumerate}
\end{thm}
\bigskip

 Our point of departure in this paper will be a periodic Schr\"odinger operator, $-\Delta+V(\bx)$, where $V$ is a honeycomb lattice potential. We fix a Dirac point, 
 ensured to exist by Theorem \ref{main-thm}, and  study the large time dynamics of wave-packets which are initially (for $t=0$) spectrally localized near these points.   Thus we have  two band dispersion surfaces, $\bk\in\brill\mapsto\mu_{b_1}(\bk)\equiv\mu_-(\bk)$ and $\bk\mapsto\mu_{b_1+1}(\bk)\equiv\mu_+(\bk)$ which touch conically at $\bk=\bK_\star$ with $\mu_\pm(\bK_\star)=\mu_\star$. 
  
  Let 
\begin{align}
 \Phi_1(\bx)\ &=_{\rm def}\ \Phi_+(\bx)\ \in L^2_{\bK_\star,\tau}\nn\\
 \Phi_2(\bx)\ &=\ \overline{\Phi_1(-\bx)}\ =_{\rm def}\ \Phi_-(\bx) \in L^2_{\bK_\star,\bar\tau}
\label{upmKstar}\end{align}
span the two-dimensional subspace of the degenerate eigenvalue $\mu_\star$ for  $H$:
\begin{align}
 H\Phi_j(\bx)\ &\equiv\ \left(-\Delta+V\right)\Phi_j=\ \mu_\star\Phi_j
 \label{Phi-j-ev-eqn}\\
 \Phi_j(\bx+\bv)\ &=\ e^{i\bK_\star\cdot\bv}\  \Phi_j(\bx),\ \ \bx\in\mathbb{R}^2,\ \ \ \bv\in\Lambda_h
 \end{align}
 We also define the periodic vectors:
 \begin{equation}
 p_1(\bx) = e^{-i\bK_\star\cdot\bx}\Phi_1(\bx),\ \ p_2(\bx) = e^{-i\bK_\star\cdot\bx}\Phi_2(\bx)\ .
 \label{phi1-phi2}
 \end{equation}
 We choose these states to be orthonormal
 \[ \left\langle \Phi_l,\Phi_m\right\rangle_{L^2(\Omega)}\ =\ \delta_{lm}\ ,\ \ \  l,m=1,2\ .\]
\bigskip

To study the time evolution \eqref{td-schroedinger} we expand the solution of the initial value problem with data $\psi_0$ using the complete set of Floquet-Bloch modes:
\begin{equation}
e^{-iHt}\ \psi_0\ =\ \sum_{b\ge1}\ \int_{\mathcal{B}_h}\ e^{-i\mu_b(\bk) t}\ 
\langle \Phi_b(\cdot;\bk),\psi_0(\cdot)\rangle\ \Phi_b(\bx;\bk)\ d\bk
\label{propagator1}
\end{equation}
Suppose $H$ has a Dirac point, $\bK$. We remark that in \eqref{propagator1} we choose the Brillouin zone, $\brill_h=\brill+\bK$, which is centered at $\bK$; see Remark \ref{recenterBZ}.
 For initial conditions which are spectrally supported near $\bK$, the time evolution depends on the precise behavior of the Floquet-Bloch modes $\Phi_\pm(\bx;\bk)$ for $\bk$ near $\bK$. While Theorem \ref{main-thm} shows that the eigenvalues $\mu_\pm(\bk)$ are Lipschitz functions in a small neighborhood of $\bk=\bK$, the eigenfunctions $\Phi_\pm(\bx;\bk)$ are not continuous functions of $\bk$ in a neighborhood of $\bK$.
\medskip

\begin{thm}\label{p-near-kD}
Let $\bK$ denote a Dirac point in the sense of Definition \ref{Diracpt-def}. In particular, let $(\Phi_\pm(\bx;\bk), \mu_\pm(\bk))$ denote the $\bk-$ pseudo-periodic eigenpairs as in part 4 of Definition \ref{Diracpt-def}. Introduce the $\Lambda_h-$ periodic functions $p_\pm(\bx;\bk)$ by
\begin{equation}
\Phi_\pm(\bx;\bk)\ =\ e^{i\bk\cdot\bx}\ p_\pm(\bx;\bk),\ \ \left\langle p_a(\cdot;\bk),p_b(\cdot;\bk)\right\rangle=\delta_{ab},\ a,b\in\{+,-\}\ .
\label{pdef}
\end{equation}
Let $\bk=\bK+\bkappa$. Then, for $0<|\bkappa|<\delta$  and $p_j$ given by 
 \eqref{phi1-phi2} we have:
 \begin{align}
 \mu_\pm(\bK+\kappa)\ &=\ \mu_\star\ \pm\ |\lambda_\sharp|\ (\kappa_1^2+\kappa_2^2)^{1\over2}\ +\ \mathcal{O}\Big(\kappa_1^2+\kappa_2^2\Big),\ \label{mu-pm-expanded}
 \end{align}
 and $p_\pm(\bx;\bk)$, {\it a priori} defined up to an arbitrary (complex) multiplicative constant of absolute value $1$, can be chosen so that:
 \begin{align} 
 p_\pm(\bx;\bK+\bkappa)
 \ &=\ \frac{1}{\sqrt{2}}\ \frac{\kappa_1+i\kappa_2}{(\kappa_1^2+\kappa_2^2)^{1\over2}}\ p_1(\bx)\ \pm\ \frac{1}{\sqrt{2}}\ p_2(\bx)\nn\\
 &\qquad\qquad +\ \mathcal{O}_{H^2(\R^2/\Lambda_h)}\Big( (\kappa_1^2+\kappa_2^2)^{1\over2}\Big)\ .
 \label{p_pm-expanded}
 \end{align}
\end{thm}
\bigskip

\nit {\it Proof of Theorem \ref{p-near-kD}:}\ The proof builds on the proof of Theorem 4.1 of \cite{FW:12}, in which the $\bk-$ pseudo-periodic eigenvalues, $\mu_\pm(\bk)$, are constructed.  These were shown to be Lipschitz continuous functions for $\bk$ varying in a neighborhood of $\bK$. We now consider the associated Floquet-Bloch modes. 

Let $\bk-$ pseudo-periodic Floquet-Bloch modes can be expressed in the form 
\[\Phi(\bx;\bk)=e^{i\bk\cdot\bx}p(\bx;\bk),\]
 where  $p(\bx;\bk)$ is $\Lambda_h$ periodic. Since we are interested in the character of $p(\bx;\bk)$ for $\bk$ near $\bK$ we set: $\bk=\bK+\bkappa$. Then, $p(\bx;\bK+\kappa)$ satisfies the periodic eigenvalue problem:
\begin{align}
& H(\bK+\bkappa)\ p(\bx;\bK+\bkappa)\ =\ \mu(\bK+\bkappa)\ p(\bx;\bK+\bkappa)\ ,\label{Hk-evp}\\
&p(\bx+\bv;\bK+\bkappa)=p(\bx;\bK+\bkappa),\ \ \textrm{for all}\ \bv\in \Lambda_h\ ,
\label{psi-per}\end{align}
where 
\[H(\bk)\ \equiv\  -\left(\nabla_\bx + i\bk\right)^2\ +\ V(\bx)\ .\]
The eigenvalue problem \eqref{Hk-evp}-\eqref{psi-per} has eigenvalues, computed via degenerate perturbation theory of the double eigenvalue $\mu_\star$ of $H(\bK)$, given by:
\begin{align}
\mu_\pm(\bK+\bkappa) &=\ \mu_\star\ +\ \mu^{(1)}\ 
 \label{mu-exp}\\
 \mu^{(1)}\ &=\ \pm|\lambda_\sharp|\ |\kappa|\ +\ \mathcal{O}(|\bkappa|^2) \ .\label{mu1-def}
  \end{align}
 
 Denote by $Q_\perp$ the  projection onto the orthogonal complement of ${\rm span}\{p_1,p_2\}$.
 Then,
 \[ R_\bK(\mu_\star)\ \equiv\ \left(\ H(\bK)\ - \mu_\star\ I \right)^{-1}:\ Q_\perp L^2(\R^2/\Lambda_h)\to Q_\perp L^2(\R^2/\Lambda_h)\]
 is bounded.
 Furthermore, via Lyapunov-Schmidt reduction of the eigenvalue problem 
\eqref{Hk-evp}-\eqref{psi-per} we obtain, corresponding to the eigenvalues $\mu_\pm(\bK+\bkappa)$ the Floquet-Bloch modes:
\begin{align}
 p_\pm(\bx;\bK+\bkappa)\ 
   &=\  
\Big(\ I\ +\ 
R_\bK(\mu_\star)\ Q_\perp\ 
\left(2i\bkappa\cdot\left(\nabla+i\bK\right)\right)\  \Big)
\left(\ \alpha p_1(\bx)\ +\ \beta p_2(\bx)\ \right)\nn\\
&\qquad\qquad + 
 \mathcal{O}_{H^2(\R^2/\Lambda_h)}\left(|\kappa|(|\alpha|^2+|\beta|^2)^{1\over2}\right)\ .\label{mu-psi-exp}
\end{align}

The $\bkappa-$ dependent coefficients  $\alpha$ and $\beta$ satisfy the homogeneous equation
\begin{equation}
 \mathcal{M}(\mu^{(1)},\kappa)\  \left(\begin{array}{c} \alpha \\ { }\\ \beta\end{array}\right)\ =\ 0\ ,
 \label{Mveca}
 \end{equation}
where  $\mathcal{M}(\mu^{(1)},\kappa)$ is a $2\times2$ matrix of the form:
{\footnotesize{
\begin{align}
\mathcal{M}(\mu^{(1)},\kappa)&\equiv\  
\left(\begin{array}{cc} 
\mu^{(1)} + \mathcal{O}\left(|\kappa|^2\right)&  
-\overline{\lambda_\sharp}(\kappa_1+i\kappa_2)+ \mathcal{O}\left(|\kappa|^2\right)  \\
&\nn\\
-\lambda_\sharp(\kappa_1-i\kappa_2) +
 \mathcal{O}\left(|\kappa|^2\right) 
 &
 \mu^{(1)}  + \mathcal{O}\left(|\kappa|^2\right)
 \end{array}\right) \ \ ;
 \end{align}\
 }}
 see \cite{FW:12} for the derivation of \eqref{mu-psi-exp}-\eqref{Mveca}.
 
 The Floquet-Bloch modes, $p_\pm(\bx;\bk)$, are finally determined by the solutions  $\alpha=\alpha_\pm(\bkappa)$, $\beta=\beta_\pm(\bkappa))$ of the homogeneous algebraic system \eqref{Mveca} for the choices $\mu^{(1)}_\pm(\bkappa)$ in \eqref{mu1-def}. We select (normalized) solutions of \eqref{Mveca} as follows:
 \begin{align}
 \mu^{(1)}_\pm &= \pm|\lambda_\sharp|\ |\bkappa| + \mathcal{O}(|\bkappa|^2),\nn\\
 \left(\begin{array}{c} \alpha_\pm \\ \beta_\pm\end{array}\right)\ &=\ 
 \left(\begin{array}{c} \frac{1}{\sqrt{2}}\ \frac{\overline{\lambda_\sharp}}{|\lambda_\sharp|}\ 
 \frac{\kappa_1+i\kappa_2}{(\kappa_1^2+\kappa_2^2)^{1\over2}} + 
 \mathcal{O}\left((\kappa_1^2+\kappa_2^2)^{1\over2}\right)\\
 \pm \frac{1}{\sqrt{2}}\ +\ \mathcal{O}\left((\kappa_1^2+\kappa_2^2)^{1\over2}\right)
 \end{array}\right)\ .
\label{mu-p}
 \end{align}
 Here, we take advantage of the observation that for a rank-1 matrix
 \[\mathcal{A}\ =\ \left(\begin{array}{cc} A & B\\ C & D\end{array}\right)\ \textrm{the vector}
  \left(\begin{array}{c} B\\ -A\end{array}\right)\]
lies in the nullspace of $\mathcal{A}$.\medskip

 Thus, with $\bkappa \equiv \bk-\bK$, we have upon substitution of \eqref{mu-p}
  into \eqref{mu-psi-exp}  the expansions of $\mu_\pm(\bk)$ and $p_\pm(\bx;\bk)$ in \eqref{mu-pm-expanded} and \eqref{p_pm-expanded}.
 \bigskip
 
\section{2D  Dirac equation}\label{2D-Dirac-eqn}

In this section we collect results on well-posedness and estimates on solutions of the two-dimensional Dirac system \eqref{Dirac-1}-\eqref{Dirac-2}.
Taking the Fourier transform of \eqref{Dirac-1}-\eqref{Dirac-2} we obtain for $\alpha(\Xi,T)=\alpha(\xi_1,\xi_2,T)$ the equation 
\begin{align}
i\D_T\left(\begin{array}{c}\hat\alpha_1 \\ \hat\alpha_2\end{array}\right)\
 &=\ \Omega(\Xi)\ 
      \left(\begin{array}{c}\hat\alpha_1 \\ \hat\alpha_2\end{array}\right),\ \ \textrm{where}\nn\\
  \Omega(\Xi)\ &=\ \left(\begin{array}{cc} 0 & \overline{\lambda_\sharp}(\xi_1+i\xi_2)\\
              \lambda_\sharp(\xi_1-i\xi_2)  & 0 \end{array}\right)\ =\ \Omega(\Xi)^*\ .
  \label{OmegaXi}    \end{align}

\begin{remark}\label{massless}
It follows from \eqref{OmegaXi} that the dispersion relation for \eqref{Dirac-1}-\eqref{Dirac-2} is:
\[\omega^2(\xi_1,\xi_2)= |\lambda_\sharp|^2\ \left(\xi_1^2+\xi_2^2\right)\ ;
\]
the effective dynamics are non-dispersive.
\end{remark}

\begin{remark}\label{dirac-structure}
The system \eqref{Dirac-1}-\eqref{Dirac-2} has the structure of Dirac system:
%
%
\begin{equation}
i\D_t\left( \begin{array}{c} \alpha_1 \\ \alpha_2 \end{array} \right)
\ =\ \left[\ \sigma_1^\sharp\ \frac{1}{i}\frac{\D}{\D X_1}\ -\ \sigma_2^\sharp\ \frac{1}{i}\frac{\D}{\D X_2}\ \right]\
\left( \begin{array}{c} \alpha_1 \\ \alpha_2 \end{array} \right)\ .
\label{dirac-system-1}
\end{equation}
The $2\times2$ matrices $\sigma_1^\sharp$ and $\sigma_2^\sharp$ satisfy the relations
\[ (\sigma_j^\sharp)^2\ =\ |\lambda_\sharp|^2\ Id,\ \ {\rm and}\ \ \sigma_1^\sharp\sigma_2^\sharp\ +\ \sigma_2^\sharp\sigma_1^\sharp\ =\ 0\ .\]
Here, $\sigma_j^\sharp,\ j=1,2$ are given by the Hermitean matrices:
\begin{align}
\sigma_1^\sharp\ \equiv\ \Lambda_\sharp\sigma_1,\ \ \sigma_2^\sharp\ \equiv\ \Lambda_\sharp\sigma_2\ ,
\label{sigma-sharp}
\end{align}
in terms of $\Lambda_\sharp$ and the standard Pauli matrices $\sigma_1$ and $\sigma_2$\ :
\begin{align}
&\Lambda_\sharp\ =\ \left(\begin{array}{cc} \overline{\lambda_\sharp} & 0 \\ 0 & \lambda_\sharp 
\end{array}\right),\ \ \sigma_1\ =\ \left(\begin{array}{cc} 0 & 1 \\ 1 & 0
\end{array}\right),\ \ \sigma_2\ =\ \left(\begin{array}{cc} 0 & -i \\ i & 0
\end{array}\right)
\end{align}
Finally, note that  $\alpha_j,\ j=1,2$ satisfy the two-dimensional wave equation, with wave-speed $|\lambda_\sharp|$:
\begin{align*}
\D^2_t\alpha\ 
  &= |\lambda_\sharp|^2\left(\frac{\D^2}{\D X_1^2}+\frac{\D^2}{\D X_2^2}\right)\alpha\ .
 \end{align*}

\end{remark}

\begin{proposition}\label{prop:dirac}\smallskip
\quad

\nit  Assume $\alpha(\bX,0)=\alpha_{0}(\bX)\in\left(H^s(\R^2)\right)^2$. Then, 
\[ \D_T^k\alpha\in L^\infty\left([0,\infty)\ ;\ \left(H^{s-k}(\R^2)\right)^2\right),\ \textrm{for}\  0\le k\le s\ .\]
 In particular,
\begin{enumerate}
\item The Fourier transform of the solution, $\alpha(\bX,T)$ is given explicitly by
 \[\hat\alpha(\Xi,T)\ =\ e^{-i\Omega(\Xi)T}\ \hat\alpha_0(\Xi),\ \ \Omega(\Xi)^*=\Omega(\Xi)\]
\item For all $\Xi\in\R^2$,\  $\left| \hat\alpha(\Xi,T) \right|\ =\ \left| \hat\alpha_0(\Xi)\right|$ and therefore 
\item For any ${\bf a}\in\Z^2$ with $|{\bf a}|\le s$ 
\begin{equation}
\left\|\ \D^{\bf a}_\bX\alpha(\bX,T)\ \right\|_{L^2(\R^2)}\ =\ \ \left\|\ \D^{\bf a}_\bX\alpha(\bX,0)\ \right\|_{L^2(\R^2)}
\label{L2-conserved}
\end{equation}
\end{enumerate}
\end{proposition}
\bigskip\bigskip

 \section{Effective Dirac dynamics; statement of main result, Theorem \ref{effective-Dirac}}\label{sec:wave-packet}

A general solution of the time-dependent Schr\"odinger equation, constrained to the  degenerate $2$-dimensional eigenspace associated with eigenvalue, $\mu_\star=\mu(\bK_\star)$, associated with the Dirac point,  $\bK_\star$,   is of the form
\begin{equation}
\psi(\bx,t)=e^{-i\mu_\star t}\ \left(\ \alpha_1\ \Phi_1(\bx) + \alpha_2\ \Phi_2(\bx)\ \right),
 \label{espace-evolve}
 \end{equation}
 where $\alpha_1$ and $\alpha_2$ are arbitrary constants.

Consider now a  {\it wave packet} initial condition, which is spectrally concentrated near $\bK_\star$:
\begin{align}
\psi^\delta(\bx,0)\ =\ \psi_0^\delta(\bx)\ &=\ \delta\ \left(\ \alpha_{10}(\delta\bx)\ \Phi_1(\bx) + \alpha_{20}(\delta\bx)\ \Phi_2(\bx)\ \right)\nn\\
&=\delta\ \left(\ \alpha_{10}(\delta\bx)\ p_1(\bx) + \alpha_{20}(\delta\bx)\ p_2(\bx)\ \right)\ e^{i\bK_\star\cdot\bx}\
 \label{packet-data}
 \end{align}
 Here, $\delta$ is a small parameter. We assume $\alpha_{10}(\bX)$ and $\alpha_{20}(\bX)$ are Schwartz functions of $\bX$. We expect that this assumption can be weakend considerably without difficulty. The overall factor of $\delta$ in \eqref{packet-data}  is not essential (the problem is linear), but is inserted so that $\psi^\delta_0$
  has $L^2(\R^2)$- norm of order of magnitude one.

We seek solutions of \eqref{td-schroedinger}, \eqref{packet-data} in the form:
\begin{equation}
\psi^\delta(\bx,t)=e^{-i\mu_\star t}\left( \sum_{j=1}^2\ \delta\ \alpha_j(\delta\bx,\delta t)\Phi_j(\bx) + \eta^\delta(\bx,t)\right).
 \label{packet-ansatz}
 \end{equation}
\begin{equation}
\textrm{where}\ \eta^\delta(\bx,0)=0,\ \alpha_j(\bX,0)=\alpha_{j0}(\bX),\ j=1,2 
\label{eta0}
\end{equation} 
to ensure the initial condition \eqref{packet-data}. 
\bigskip

The goal is to show 
that the Schr\"odinger equation \eqref{td-schroedinger} has a solution of the form \eqref{packet-ansatz} with an error term, $\eta^\delta(\bx,t)$, which satisfies
\begin{equation}
\sup_{0\le t\le \rho\delta^{-2+\eps_1} }\ \|\ \eta^\delta(\cdot,t)\ \|_{H^s(\R^2)}\ =\ 
\mathcal{O}(\delta^{\tau_\star}),\ \ \delta\to0\ .
\label{eta-goal}
\end{equation}
for some $\tau_\star>0$, 
  provided  the slowly varying amplitudes $\alpha_j(\delta\bx,\delta t),\ j=1,2$ evolve according to the system of Dirac-type equations \eqref{Dirac-1}-\eqref{Dirac-2}.   Here, $\rho>0$ and $\eps_1>0$ are arbitrary.
\medskip

 We shall prove the following
\medskip
 
 \begin{theorem}\label{effective-Dirac} Assume 
 \[ \vec{\alpha}_0(\bX)\ \equiv\ \left(\begin{array}{c}\alpha_{10}(\bX)\\ 
 \alpha_{20}(\bX)\end{array}\right) \in \left[\mathcal{S}(\R^2)\right]^2\]
 and let $\vec{\alpha}(\bX,T)$ denote the global-in-time solution of the Dirac system \eqref{Dirac-1}-\eqref{Dirac-2}
 %
 with initial data $\vec\alpha(\bX,0)=\vec\alpha_0(\bX)$. 
 Consider the time-dependent Schr\"odinger equation, \eqref{td-schroedinger}, where $V(\bx)$ denotes a potential for which the conclusions of Theorem \ref{main-thm} hold, {\it e.g.} 
 $V(\bx)=\epsilon V_h(\bx)$, where $V_h$ is a honeycomb lattice potential satisfying $V_{1,1}\ne0$ and  $\epsilon$ is not in the countable closed set $\tilde{\mathcal{C}}$.  Assume initial conditions, $\psi_0$,  of the form
  \eqref{packet-data}. 
 Fix any $\rho>0$ and $\eps_1>0$. Also choose non-negative $N=(n_1,n_2)\in\mathbb{Z}^2$ with $n_1,n_2\ge0$.  Then, \eqref{td-schroedinger} has a unique solution of the form \eqref{packet-ansatz}, where for any $|\alpha|\le N$
 \begin{equation}
 \sup_{ 0\le t\le \rho\ \delta^{-2+\eps_1} } 
 \left\|\ \D_{\bx}^\alpha
  \eta^\delta(\bx,t)\ \right\|_{L^2(\R_\bx^2)}\  \ =\ o(\delta^{\tau_\star}),\ \ \delta\to0\,
  \label{eta-estimate}
 \end{equation}
 for some $\tau_*>0$. 
 \end{theorem}
 
 \section{Proof of Theorem \ref{effective-Dirac}}\label{pf-effectiveD}

We begin with a  summary of  consequences of
Theorems \ref{main-thm} and \ref{p-near-kD}  which  we use 
in the proof.
\medskip

Recall that the spectral bands $\mu_+(\brill_h)$ and $\mu_-(\brill_h)$ touch conically at the the vertices of $\brill_h$. Specifically, 
 $\mu_+(\bk)=\mu_-(\bk)=\mu_\star$ for $\bk=\bK_\star$ equal to any of the three $\bK$- type vertices, $\{\bK,R\bK,R^2\bK\}$ and any of the three $\bK'$- type vertices,  $\{\bK',R\bK',R^2\bK'\}$,  where $\bK'=-\bK$. Here, $R$ denotes the $\pi/3$- counterclockwise rotation. 
 \medskip
 
\nit For a Dirac point $\bK_\star$, equal to any vertex of $\brill_h$:
 \begin{itemize}
 \item[(P1)] There exist $L^2_\bk$- eigenvalues of $H$, denoted $\mu_\pm(\bk)$, such that  
  \begin{align}
&  \textrm{for all}\  \bk\in\mathcal{B}_h\ \textrm{ satisfying}\ |\bk-\bK_\star|<\kappa_0\nn\\
&  \mu_\pm(\bk)-\mu(\bK_\star)=
\pm |\lambda_\sharp|\left|\bk-\bK_\star\right|\left(1+E_\pm(\bk-\bK_\star)\right),
\label{mupm-expand} \end{align}
where $E_\pm(\bkappa)$ are  Lipschitz continuous in $\bkappa$ and 
$E_\pm(\bkappa)\le C_\pm\ |\bkappa|$.
 Here, $\lambda_\sharp$ is a constant given by \eqref{lambda-sharp2}.
 %
 %
\item[(P2)] There are constants $\kappa_1, C_1$ such that
\begin{align}
&\textrm{ for all}\  \bk\in\mathcal{B}_h\ \textrm{satisfying}\ 
|\bk-\bK_\star|<\bkappa_1\ \textrm{ and all}\ b\notin\{+,-\} \nn\\ 
 & \left|\mu_b(\bk)-\mu(\bK_\star)\right|\ge C_1>0.
 \label{mupm-lowerbd}
 \end{align}
  \item[(P3)]   For $0<|\bkappa|<\delta$ sufficiently small
   \begin{align}
p_\pm(\bx;\bK_\star+\bkappa)\ &=\
 \frac{1}{\sqrt{2}}\Big(\ \alpha(\bkappa)\ p_1(\bx)\ \pm\ p_2(\bx)\ \Big)\ +\ 
 \mathcal{O}_{H^2(\R^2/\Lambda_h)}\left(|\bkappa|\right),\label{p_pm-locally1}\\
\ {\rm where}\ \ \alpha(\bkappa)\ &=\ \frac{\overline{\lambda_\sharp}}{|\lambda_\sharp|}\ \frac{\bkappa_1+i\bkappa_2}{\sqrt{\bkappa_1^2+\bkappa_2^2}}
\label{alpha-def1}
\end{align}
%
%
\end{itemize}
\nit (P2) is a consequence of (P1) and the continuity of the eigenvalues $\bk\mapsto\mu_b(\bk)$;
 see Proposition \ref{ktomuk}.
\bigskip

\nit {\it Without loss of generality and for simplicity:}
\begin{enumerate}
\item we take
\begin{align}
& \bK_\star=\bK
\nn\end{align}

\item we recall that the  Brillouin zone, $\brill_h$, is centered at $\bK$; see Remark \ref{recenterBZ}. 
Since we assume wave-packet initial data which are spectrally localized near $\bK$ in $\R^2/\Lambda^*_h$, 
this equivalent choice, which puts $\bK$ on the interior of the Brillouin zone, simplifies the analysis. 
\end{enumerate}
 To prove Theorem \ref{effective-Dirac}, we study the evolution equation for $\eta^\delta(\bx,t)$  obtained by substitution of \eqref{packet-ansatz} into \eqref{td-schroedinger}: 
\begin{align}
&i\D_t\eta^\delta\ =\ \left(H-\mu_\star\right)\eta^\delta\nn\\
&\qquad\  -\delta^2  \sum_{j=1}^2\ \left.\Big[\ i\D_T\alpha_j(\bX,T)\ \Phi_j(\bx)+2\nabla_\bX\alpha_j(\bX,T)\cdot\nabla_\bx\Phi_j(\bx)\Big]\ \right|_{(\bX,T)=(\delta\bx,\delta t)} \nn\\
&\qquad\ -\ \delta^3\ \sum_{j=1}^2\ \Delta_\bX\alpha_j(\bX,T)\Phi_j(\bx)\ \Big|_{(\bX,T)=(\delta\bx,\delta t)},\label{eta-eqn}\\
&\eta^\delta(\bx,t=0)\ =\ 0\ .\label{eta-data}
\end{align}

\subsection{Estimation of the error, $\eta^\delta(\bx,t)$}
 
 Using the DuHamel principle we may rewrite \eqref{eta-eqn} as the equivalent integral equation:
 {\small{
 \begin{align}
 \eta^\delta(\cdot,t)\ &=\ 
  i\delta^2\  \sum_{j=1}^2\ \int_0^t\ e^{-i(H-\mu_\star)(t-s)}\  
 \Big[\ i\D_S\alpha_j(\delta\cdot,\delta s)\ \Phi_j(\cdot)+2\nabla_\bX\alpha_j(\delta\cdot,\delta s)\cdot\nabla_x\Phi_j(\cdot)\ \Big]\ ds \nn\\
&\qquad\ +\ i\delta^3\  \sum_{j=1}^2\int_0^t\ e^{-i(H-\mu_\star)(t-s)}\ \Delta_\bX\alpha_j(\delta\cdot,\delta s)\Phi_j(\cdot)\  ds\label{eta-integ-eqn}
\end{align}
}}

The second integral in \eqref{eta-integ-eqn} (call it $Int_2$) can be bounded as follows.  
 Let $s$ denote an even positive integer. Recall from elliptic theory that
 \begin{equation}
  \|f\|^2_{H^s}\ \lesssim\  \|f\|_2^2\ +\ \|(-\Delta+V+c)^{s\over2}f\|^2_2\ \lesssim \|f\|^2_{H^s}\ ,
  \label{Hs-equiv}
  \end{equation}
  where $c>\sup_{\bx\in\R^2}|V(\bx)|$. 
  Using \eqref{Hs-equiv},  we can bound $\|Int_2\|_{H^s}$  in terms of  $\|Int_2\|_{L^2}$ and 
  $\|(-\Delta+V+c)^{s\over2}Int_2\|_{L^2}$. Taking the $L^2$ norm of $Int_2$ and using that $e^{-iHt}$ is unitary in $L^2$, we obtain 
  \begin{equation}
   \|Int_2\|_{L^2}\lesssim \delta^3\sum_{j=1}^2\int_0^t\|\Delta_\bX\alpha_j(\delta\cdot,\delta s)\|_{L^2}\ ds.\label{Int2-L2}
   \end{equation}
   Next, using that 
  $H=-\Delta+V$ commutes with $e^{-iHt}$ we obtain 
 \begin{equation}
  \|(-\Delta+V+c)^{s\over2}Int_2\|_{L^2}\lesssim\delta^3\sum_{j=1}^2
 \int_0^t\ \|(-\Delta+V+c)^{s\over2}\Delta_\bX\alpha_j(\delta\cdot,\delta s)\|_{L^2}\ ds. 
 \label{DeltaInt2-L2}
 \end{equation}
Next note, via Proposition \ref{prop:dirac}, that
 \begin{equation}
  \|\alpha_j(\delta \cdot,\delta t)\|_{H^{m}(\R^2)}\le C\delta^{-1}\ \|\alpha_0\|_{H^{m}(\R^2)}\ .
  \label{Hm-bound}
  \end{equation}
 Let $\eps_1>0$ be arbitrary. Then, equations \eqref{Hs-equiv}-\eqref{Hm-bound} imply 
 {\small{
 \begin{align}
&\sup_{0\le t\le \rho \delta^{-2+\eps_1}}\|Int_2\|_{H^s}\nn\\
& \equiv\ \sup_{0\le t\le \rho \delta^{-2+\eps_1}}\ \left\|\ \delta^3\ \sum_{j=1}^2 \int_0^t\ e^{i(H-\mu_\star)(t-s)}\ \Delta_\bX\alpha_j(\delta\cdot,\delta s)\Phi_j(\cdot)\  ds\ \right\|_{H^s(\R^2)}\ \le\ C\ \rho\ \delta^{\eps_1}\ .
\label{last-term-bound}
\end{align}
}}

It therefore suffices to estimate the first time-integral in \eqref{eta-integ-eqn}.
 This time-integral is the solution of the initial value problem of the form:
 \begin{align}
 &i\D_t f^\delta(\bx,t)\ -\ \left(H-\mu_\star\right) f^\delta(\bx,t)\ =\ 
 \delta^2\ \sum_r \gamma_r(\delta\bx,\delta t)\Psi_r(\bx),
 \label{eta-form}\\
 & f^\delta(\bx,0)=0, 
 \end{align}
where we find it convenient to introduce the notation
 \begin{equation}
 \sum_r \gamma_r(\bX,T)\Psi_r(\bx) \ \equiv\
 -\sum_{j=1}^2\ \Big[\  i\D_T\alpha_j(\bX,T)\ \Phi_j(\bx)
 \ +\ 2 \nabla_\bX\alpha_j(\bX,T)\cdot \nabla_\bx\Phi_j(\bx) \ \Big]\ ,   
\label{gammaPsi-def}   \end{equation}
where
\begin{align}
\gamma_1(\bX,T)\ &=\ i\D_T\alpha_1(\bX,T),\qquad \Psi_1(\bx)=\Phi_1(\bx),\label{g1}\\
\gamma_2(\bX,T)\ &=\ i\D_T\alpha_2(\bX,T),\qquad \Psi_2(\bx)=\Phi_2(\bx),\label{g2}\\
\gamma_3(\bX,T)\ &=\ 2\nabla_\bX\alpha_1(\bX,T),\qquad \Psi_3(\bx)=
(\Psi_{3,1}(\bx),\Psi_{3,2}(\bx))=\nabla_\bx\Phi_1(\bx),\label{g3}\\
\gamma_4(\bX,T)\ &=\ 2\nabla_\bX\alpha_2(\bX,T),\qquad \Psi_4(\bx)=
(\Psi_{4,1}(\bx),\Psi_{4,2}(\bx))=\nabla_\bx\Phi_2(\bx)\ .\label{g4}
\end{align}
Note that $\Psi_j(\bx)\in L^2_\bK,\ 1\le j\le4$.


By hypotheses on $\alpha_j$ and $\Phi_j$, and Proposition \ref{prop:dirac}, the functions $\gamma_r$ and $\Psi_r$ satisfy the following properties:
 \begin{align}
 &\gamma_r(\bX,T)\ \textrm{is}\ C^\infty\left(\R_T;\mathcal{S}(\R_{\bf X}^2)\right)\ 
\label{gamma-r}\\
& \| \gamma_r(\cdot,T)\|_{H^s(\R^2)}\ \lesssim\  \| \alpha(\cdot,T=0)\|_{H^{s+1}}\label{gammaHs}\\
 &\Psi_r(\bx)\in C^\infty(\R_\bx^2)\cap L^2_{\bK}\nn\\
 &  \textrm{and therefore satisfies the pseudo-periodic boundary condition}\nn\\ 
 & \Psi_r(\bx+\bv)\ =\ e^{i\bK\cdot\bv}\Psi_r(\bx)\ .\label{Psi-r-pseudo}
  \end{align}
 We also write $\Psi_r$  in the following useful form:
   \begin{equation}\label{psi-form}
  \Psi_r(\bx)\ =\ e^{i\bK \cdot\bx}\mathcal{P}_r(\bx),\ \ \mathcal{P}_r(\bx+\bv)=\mathcal{P}_r(\bx)\ ,\ \bv\in\Lambda_h\ .
  \end{equation}
  By \eqref{Dpb}, for any $\beta\in\mathbb{N}_0^2$,  
\begin{equation}
\| \D^\beta_\bx\mathcal{P}_r \|_{L^\infty(\Omega)}\ \le\ C_{r,\beta}\ .
\label{DPr}
\end{equation}
  
  Theorem \ref{effective-Dirac} is therefore reduced to the following \medskip
  
  \begin{proposition}\label{average-fast}
  Let $\alpha_1(\bX,T)$ and $\alpha_2(\bX,T)$ satisfy the system of Dirac equations:
  \begin{align}
\D_T\alpha_1\  =\ -\overline{\lambda_\sharp}\left(\D_{X_1}+i\D_{X_2}\right)\alpha_2\label{alpha1eqn}\\
\D_T\alpha_2\ =\   -\lambda_\sharp\left(\D_{X_1}-i\D_{X_2}\right)\alpha_1\label{alpha2eqn}
\end{align}
with initial conditions as in Theorem \ref{effective-Dirac}.
  Then, for any $s\ge1, \rho>0$ and $\eps_1>0$,  there exists a unique solution, $f^\delta(\bx,t)$ of \eqref{eta-form}, which satisfies the estimate 
   \begin{equation}
 \sup_{ 0\le t\le \rho\ \delta^{-2+\eps_1} }\ 
 \left\|\ 
  f^\delta(\cdot,t)\ \right\|_{H^s(\R^2)}\ \le\ C\ \delta^{\frac{\eps_1}{2}},\ {\rm for}\  \delta\downarrow0\ .
    \label{f-estimate}
 \end{equation}
  \end{proposition}
  \medskip
  
\bigskip
  So to finish the proof Theorem \ref{effective-Dirac}, we only require a proof of  Proposition \ref{average-fast}. 

\section{Proof of Proposition  \ref{average-fast}}
\label{proof-of-fast-avg}

By the completeness of Bloch modes,
\begin{equation}
e^{-i(H-\mu_\star)t}\ g\ =\ \sum_{b}\ \int_{\mathcal{B}_h}\ e^{-i(\mu_b(\bk)-\mu_\star)t}\ 
\langle \Phi_b(\cdot;\bk),g(\cdot)\rangle_{L^2(\R^2)}\ \Phi_b(\bx;\bk)\ d\bk
\label{propagator}
\end{equation}
Thus,
\begin{equation}
f^\delta(\bx,t) = \sum_{b}\ \int_{\mathcal{B}_h}\ \tilde f^\delta_b(\bk,t) \Phi_b(\bx;\bk)\ d\bk,
\label{eta-sum}
\end{equation}
where
\begin{equation}
\tilde f^\delta_b(\bk,t) = -i\delta^2\ \int_0^t\ ds\ e^{-i(\mu_b(\bk)-\mu_\star)(t-s)}\ 
 \left\langle \Phi_b(\cdot;\bk) , \sum_r\gamma_r(\delta\cdot,\delta s)\Psi_r(\cdot)\right\rangle_{L^2(\R^2)} \ .
\label{teta-far} \end{equation}
 {\it We shall henceforth omit the superscript $\delta$ from $f^\delta$ and $\tilde f^\delta_b$}.\medskip
 
 Decompose $f$ into $f_D$,  frequency components which lie in the  two spectral bands: $\mu_+(\brill_h)$ and $\mu_-(\brill_h)$ intersecting at the Dirac point $\mu_\star$, and $f_{D^c}$, frequency components which lie in all other spectral bands:
 \begin{align}
f(\bx,t) &= f_D(\bx,t) + f_{D^c}(\bx,t),\ \ \ \ \ \ {\rm where}\nn\\
f_D(\bx,t)\ &=\ \sum_{b\in\{+,-\}}\ \int_{\mathcal{B}_h}\ \tilde f_b(\bk,t) \Phi_b(\bx;\bk)\ d\bk\label{eta-pm}
\end{align}
and
\begin{align}
f_{D^c}(\bx,t)\ &=\  \sum_{b\notin\{+,-\}}\ \int_{\mathcal{B}_h}\ \tilde f_b(\bk,t) \Phi_b(\bx;\bk)\ d\bk\ .
\label{eta-far}\end{align}


Let's focus initially on $f_D(\bx,t)$. We distinguish between frequencies which are ``near'' and ``bounded away from'' $\mu_\star$, as these correspond to whether the complex phase in \eqref{propagator} is non-oscillatory or, respectively, oscillatory in $t$. Recalling property (P2) and our choice of  Brillouin zone with $\bK$ in its interior: 
we  further decompose $f_D$ into its quasi-momentum components near and away from 
from any of the points $\bK$:
\begin{align}
f_D(\bx,t)\ 
&=\ \sum_{b\in\{+,-\}}\ \int_{\mathcal{B}_h}\ \chi\Big(|\bk-\bK|<\delta^\tau\Big)\ \tilde f_b(\bk,t)\ \Phi_b(\bx;\bk)\ d\bk\nn\\
& +\  \sum_{b\in\{+,-\}}\ \int_{\mathcal{B}_h}\ \chi\Big(\delta^\tau\le |\bk-\bK|\Big)\ \tilde f_b(\bk,t)\ \Phi_b(\bx;\bk)\ d\bk\nn\\
&\equiv\ \sum_{b\in\{+,-\}}\ \int_{\mathcal{B}_h}\ \left[\ \tilde f_{I,b}(\bk,t)\ + \tilde f_{II,b}(\bk,t)\ \right]\ \Phi_b(\bx;\bk)\ d\bk\nn\\
&\equiv\  \sum_{b\in\{+,-\}}\ \left(\ f_{I,b}(\bx,t)\ +\ f_{II,b}(\bx,t)\ \right)\nn\\
&\equiv\  f_{I,D}(\bx,t)\ +\ f_{II,D}(\bx,t)\  . \label{eta-pm-split}
\end{align} 
 Here, $0<\tau<1$ will  chosen less than but close to $1$.
%
%
%
\nit By \eqref{plancherel} and \eqref{Hs} 
\begin{align}
\| f(\cdot,t)\|_{H^s(\R^2)}^2\ &\approx\ 
  \| f_D(\cdot,t)\|_{L^2(\R^2)}^2\ +\ \| f_{D^c}(\cdot,t)\|_{H^s(\R^2)}^2\nn\\
 &\approx\  \sum_{J=I,II}\ \| f_{J,D}(\cdot,t)\|_{L^2(\R^2)}^2\ +\ \| f_{D^c}(\cdot,t)\|_{H^s(\R^2)}^2\nn\\
 &=\ \sum_{J=I,II}\ \sum_{b=\pm} \| f_{J,b}(\cdot,t)\|_{L^2(\R^2)}^2\ +\ 
  \| f_{D^c}(\cdot,t)\|_{H^s(\R^2)}^2,
 \label{f-norm}\end{align}
 where we have used \eqref{cor-Hs}.\ 
 We show below, for fixed $\rho>0, \eps_1>0$ that each term in \eqref{f-norm} is  $\mathcal{O}(\delta^{\eps_1})$ 
   for $0\le t\le \rho\ \delta^{-2+\eps_1}$ as $\delta\downarrow0$.\bigskip
 
 In the calculations below we shall require a detailed expansion of inner products of the form:
 \begin{equation}
 \left\langle \Phi_b(\cdot;\bk) , 
\Gamma(\delta\cdot,\delta s)\Psi(\cdot)\right\rangle_{L^2(\R^2)}\ ,
\label{ip-form}\end{equation}
where $\Gamma=\Gamma(\bX,T)$ is in Schwartz class and $\Psi\in L^2_\bK$, {\it i.e.}
\begin{equation}
\Psi(\bx)\ =\ e^{i\bK\cdot\bx}\ \mathcal{P}(\bx),\ \mathcal{P}(\bx+\bv)=\mathcal{P}(\bx),\ \bv\in\Lambda_h;
\label{PsicalP}\end{equation}
 
see  \eqref{teta-far}. The following proposition will be used: 
\medskip

 \begin{proposition}\label{p-sum-ip}
 Let $\Gamma(\bX,T)$  denote a Schwartz class function of $\bX$, varying smoothly in $T$. 
 Denote by $\widehat\Gamma(\Xi,T)$ its Fourier transform with respect to the $\bX$ variable.
  Then, 
\begin{align}
& \left\langle \Phi_b(\cdot;\bk) , 
\Gamma(\delta\cdot,\delta s)\Psi(\cdot)\right\rangle_{L^2(\R^2)}
\nn\\
&=\ \int_\Omega\ \overline{p_b(\by;\bk)}\cdot \Big[\ \delta^{-2}\ \sum_{\bm\in\Z^2}\ e^{i\bm\cdot\by}\ \widehat{\Gamma}\left(\frac{m_1\bk_1+m_2\bk_2+(\bk-\bK)}{\delta},\delta s\right)\ \Big]\  
  \mathcal{P}(\by)\ d\by
\label{poisson-ip}  \end{align}
 \end{proposition}
{\it Proof of Proposition \ref{p-sum-ip}:}\ 
Recall from \eqref{completeset} and \eqref{psi-form}  that $\Phi_b(\bx;\bk)=e^{i\bk\cdot\bx}p_b(\bx;\bk)$, where  $p_b(\bx+\bv;\bk)=p_b(\bx;\bk)$ for any $\bv\in\Lambda_h$.    Thus, using \eqref{PsicalP}, we find that 
 \begin{align*}
& \left\langle \Phi_b(\cdot;\bk) , 
\Gamma(\delta\cdot,\delta s)\Psi(\cdot)\right\rangle_{L^2(\R^2)}\nn\\
 &\ \ =\ \int_{\R^2} e^{-i(\bk-\bK)\cdot\by}\ \Gamma(\delta\by,\delta s)\ \overline{p_b(\by;\bk)}\ \mathcal{P}(\by)\ d\by\\
  &\ \ =\ \sum_{\bm\in\Z^2}\ \int_\Omega\ e^{-i(\bk-\bK)\cdot(\by+\bm\bv)}\ 
  \Gamma(\delta(\by+\bm\bv),\delta s)\ \overline{p_b(\by+\bm\bv;\bk)}\ 
  \mathcal{P}(\by+\bm\bv)\ d\by\\
  &\ \ =\ \sum_{\bm\in\Z^2}\ \int_\Omega\ e^{-i(\bk-\bK)\cdot(\by+\bm\bv)}\ 
  \Gamma(\delta(\by+\bm\bv),\delta s)\ \overline{p_b(\by;\bk)}\ 
  \mathcal{P}(\by)\ d\by\\
  &\ \ =\  \int_\Omega\ \Big[\ \sum_{\bm\in\Z^2}\ e^{-i(\bk-\bK)\cdot(\by+\bm\bv)}\ 
    \Gamma(\delta(\by+\bm\bv),\delta s)\ \Big]\ \overline{p_b(\by;\bk)}\ 
  \mathcal{P}(\by)\ d\by
 \end{align*}
The above sum  can be re-written via the Poisson summation formula 
 as
\begin{align}
& \sum_{\bm\in\Z^2}\ e^{-i(\bk-\bK)\cdot(\by+\bm\bv)}\ 
    \Gamma(\delta(\by+\bm\bv),\delta s)\nn\\
    &  =\ \delta^{-2}\ \sum_{\bm=(m_1,m_2)\in\Z^2}\ e^{i\bm\cdot\by}\ \widehat{\Gamma}\left(\frac{m_1\bk_1+m_2\bk_2+(\bk-\bK)}{\delta},\delta s\right)
 \label{p-sum}
 \end{align}
 This completes the proof of Proposition \ref{p-sum-ip}.
\medskip

 
\subsection{Estimation of $\|f_{I,D}(\cdot,t)\|_{L^2(\R^2)}$}\label{eta-I-estimate}\bigskip
{\qquad\ }\bigskip

In this section we prove that for any fixed $\rho>0$ and $\eps_1>0$, 
\begin{equation}
\sup_{0\le t\le\rho\delta^{-2+\eps_1}}\ \|f_{I,D}(\cdot,t)\|_{L^2(\R^2)}\ =\ 
\mathcal{O}(\delta^{\frac{\eps_1}{2}}),\ \ \delta\downarrow0
\label{f1-est}
\end{equation}

Since
\[
f_{I,D}(\bx,t)\ =\ \sum_{b\in\{+,-\}}\ \int_{\mathcal{B}_h}\ 
\chi\Big(|\bk-\bK|<\delta^\tau\Big)\ \tilde f_b(\bk,t)\ \Phi_b(\bx;\bk)\ d\bk\ ,
\]
 we have 
 \begin{equation}
 \|f_{I,D}(\cdot,t)\|_{L^2(\R^2)}^2\ =\ \sum_{b\in\{+,-\}}\ \int_{\brill_h}\ 
 \chi\Big(|\bk-\bK|<\delta^\tau\Big) |\tilde f_b(\bk,t)|^2\ d\bk
 \end{equation}
where, by \eqref{teta-far}
\begin{equation}
\tilde f_\pm(\bk,t) = -i\delta^2\ \int_0^t\ ds\ e^{-i(\mu_\pm(\bk)-\mu_\star)(t-s)}\ 
 \sum_{r=1}^4\left\langle \Phi_\pm(\cdot;\bk) , \gamma_r(\delta\cdot,\delta s)\Psi_r(\cdot)\right\rangle_{L^2(\R^2)}\ .
\label{eta-near-model}
 \end{equation}

We next use Proposition \ref{p-sum-ip} to re-express the inner products  appearing in \eqref{eta-near-model} as follows:
 \begin{align}
& \left\langle \Phi_\pm(\cdot;\bk) , 
\gamma_r(\delta\cdot,\delta s)\Psi_r(\cdot)\right\rangle_{L^2(\R^2)}\nn\\
&=\ \int_\Omega\ \Big[\ \delta^{-2}\ \sum_{\bm\in\Z^2}\ e^{i\bm\cdot\by}\ \hat{\gamma}_r\left(\frac{m_1\bk_1+m_2\bk_2+(\bk-\bK)}{\delta},\delta s\right)\ \Big]\ \overline{p_\pm(\by;\bk)}\ 
  \mathcal{P}_r(\by)\ d\by\label{ip-psum}
  \end{align}
  Since our goal here is to estimate $f_{I,\pm}$, we recall that $\bk$ is restricted
  to:$\{|\bk-\bK|\le\delta^\tau\}$. 
 %
 
Thus, we rewrite the sum \eqref{ip-psum} in terms of its $\bm=0$ and $\bm\ne0$ contributions: 
{\small{
  \begin{align}
&\left\langle \Phi_\pm(\cdot;\bk) , 
\gamma_r(\delta\cdot,\delta s)\Psi_r(\cdot)\right\rangle_{L^2(\R^2)}\nn\\ 
&\nn\\
&=
\underbrace{
\frac{1}{\delta^{2}} \hat{\gamma}_r\left(\frac{(\bk-\bK)}{\delta},\delta s\right) \int_\Omega\ 
 \overline{p_\pm(\by;\bk)} 
  \mathcal{P}_r(\by)\ d\by}_{{\rm Term}_{1,r}} \ +\ 
 \underbrace{ \int_\Omega\ \frac{1}{\delta^{2}} E_{r,\delta}\left(\frac{\by}{\delta},\delta s;\bk\right)\ \overline{p_\pm(\by;\bk)}\ 
  \mathcal{P}_r(\by)\ d\by}_{{\rm Term}_{2.r}}
 \label{ip-re-expressed}
\end{align}
}}
where
\begin{align}
  &\ \delta^{-2}\ E_{r,\delta}\left(\delta^{-1}\by,\delta s;\bk\right)\nn\\
  & =\  \  \delta^{-2}\ 
  \sum_{\bm\in\Z^2\setminus\{(0,0)\}}\ e^{i\bm\cdot\by}\ 
  \hat{\gamma}_r\left(\frac{m_1\bk_1+m_2\bk_2+(\bk-\bK)}{\delta},\delta s\right)\ .\nn\\
\label{Erdelta-def} \end{align}
  We now study the terms 
  ${\rm Term}_{2,r}$  and ${\rm Term}_{1,r}$ in \eqref{ip-re-expressed}\ .
\bigskip

\nit{\bf Estimation of $\textrm{Term}_{2,r}$ of \eqref{ip-re-expressed}:} We begin with the following
\medskip

\begin{proposition}\label{gammahat-est}
Denote by $\gamma_r$ any of the functions $\D_T\alpha_j,\ \D_{X_l}\alpha_j,\ j,l=1,2$. 
For $A\ge0$, assume $\alpha_0(\bX)=\alpha(\bX,0)\in W^{A+1,1}(\R^2)$.  Then, there exist positive constants $C_1, C_2$ such that the following holds:
\begin{enumerate}
\item For any $A\ge0$ and all $\bkappa\in\R^2$:
\begin{align}
 \left|\hat\gamma_r(\bkappa,T)\right|\ &\le\ \frac{C_1}{(1+|\bkappa|)^A}\   \|\alpha_0\|_{W^{A+1,1}(\R^2)},\ \label{gammahat}\
 \end{align}
 \item For all  $A>2$ and all  $\bk\in\brill_h$ such that $|\bk-\bK|<\delta^\tau$ we have
 \begin{align}
 \left|\ E_{r,\delta}(\by,T;\bk)\ \right|\ &\le\ C_2\ \delta^A\    
  \left\|\alpha_0\right\|_{W^{A+1,1}(\R^2)}
\label{Erdelta}\end{align}
\end{enumerate}
\end{proposition}
{\it Proof of Proposition \ref{gammahat-est}:}  
 Each function $\gamma_r$ itself is the component of a solution of the Dirac system. Hence, by Proposition \ref{prop:dirac} and integration by parts:
\begin{align}
\left| \hat\gamma_r(\bkappa,T) \right|\  &\le\  C\ \left| \left(\D_X\alpha\right)\hat(\bkappa,T)\right|\ =\  C\  \left| \left(\D_X\alpha_0\right)\hat(\bkappa)\right| \nn\\
 &\sim\ \Big|\ \int e^{i\bkappa\cdot\bX}\D_X\alpha_0(\bX)\ d\bX\ \Big|\ \lesssim\  \frac{1}{(1+|\bkappa|)^A}\  \|\alpha_0\|_{W^{A+1,1}(\R^2)}
 \end{align}
 This proves \eqref{gammahat}. 

We now turn our attention to the bound \eqref{Erdelta}.  Note that the sum in the definition of $E_{r,\delta}$,  \eqref{Erdelta-def}, is over $\bm=(m_1,m_2)\in\Z^2\setminus\{(0,0)\}$. For such 
$\bm$, and for $\bk$ such that $|\bk-\bK|<\delta^\tau$, we have
\[ \left| m_1\bk_1+m_2\bk_2+(\bk-\bK) \right|\ge c|\bm|\ .\]
Thus, by Proposition \ref{gammahat-est}
{\small{
\begin{align}
&\bk\in\brill_h,\ \ |\bk-\bK| < \delta^\tau\ \implies\nn\\
 &\left|\ E_{r,\delta}(\by,T;\bk)\ \right|\le\ 
\left|\ \sum_{\bm\in\Z^2\setminus\{(0,0)\}}\ e^{i\bm \cdot\by}\ 
\hat{\gamma}_r\left(\frac{m_1\bk_1+m_2\bk_2+(\bk-\bK)}{\delta},T\right)\ \right|
 \nn\\
 &\le C\  \sum_{|\bm|\ge1} \frac{1}{|\bm|^A}\ \delta^A\ 
   \left\|\alpha_0\right\|_{W^{A+1,1}(\R^2)}
 \end{align}
 }}
 where we take $A>2$ for the sum to converge. This completes the proof of  the bound \eqref{Erdelta}
  and therewith Proposition \ref{gammahat-est}. 
 \medskip
 
By Proposition \ref{gammahat-est} and \eqref{DPr} we obtain the following bound on the 
last term in \eqref{ip-re-expressed}:
 \begin{align}
\bk\in\brill_h,\ \ |\bk-\bK| &< \delta^\tau\ \implies\nn\\
\left| \textrm{Term}_{2,r}\right|\ &\equiv\ \left|\ \int_\Omega\  \delta^{-2}\ E_{r,\delta}\left(\delta^{-1}\by,\delta s;\bk\right)\ 
\overline{p_\pm(\by;\bk)}\ 
  \mathcal{P}_r(\by)\ d\by\ \right|\nn\\
  & \lesssim\ 
  \delta^{A-2}\ 
  \left\|\alpha_0\right\|_{W^{A+1,1}(\R^2)},\ \ A>2\ .
\label{int-Erdelta}  \end{align}
 \medskip
 
 \nit{\bf Expansion and estimation of $\textrm{Term}_{1,r}$ of \eqref{ip-re-expressed}:}\ \ 
$\textrm{Term}_{1,r}$ may be written as:
  \begin{align}
\textrm{Term}_{1,r}\ &=\ 
   \delta^{-2}\ \chi\left(|\bK-\bk|<\delta^\tau\right)
    \cdot \hat{\gamma}_r\left(\ \frac{\bk-\bK}{\delta},\delta s\right)\  
 \cdot \ 
 \left\langle p_\pm(\cdot;\bk),\mathcal{P}_r(\cdot) \right\rangle_{L^2(\Omega)}\ . \label{T1-rewritten}
 \end{align}
 \bigskip

Note that the argument of $\hat{\gamma}_r$ may be small and hence we may no longer use the decay properties of $\hat\gamma_r$ to control the magnitude of \eqref{T1-rewritten}. We therefore use the precise behavior of $p_\pm(\bx;\bk)$ for  $\bk$ near $\bK$. By (P3)
\begin{align}
 p_\pm(\bx;\bk)
 \ &=\ \frac{1}{\sqrt{2}}\ \frac{(\bk_1-\bK_1)+i(\bk_2-\bK_2)}{((\bk_1-\bK_1)^2+(\bk_2-\bK_2)^2)^{1\over2}}\ p_1(\bx)\ \pm\ \frac{1}{\sqrt{2}}\ p_2(\bx)\nn\\
 &\qquad\  +\ \mathcal{O}_{H^2(\R^2/\Lambda_h)}\Big( \left( (\bk_1-\bK_1)^2+(\bk_2-\bK_2)^2 \right)^{1\over2}\Big)
 \label{p-pm-expanded1}
 \end{align}
 
Next we use \eqref{p-pm-expanded1} to expand the inner product in  \eqref{T1-rewritten}. We have, recalling that $\bk=\bK+\bkappa$:
{\small{
\begin{align}
& \hat{\gamma}_r\left(\ \frac{\bk-\bK}{\delta},\delta s\right)\  
 \cdot \ 
\left\langle p_\pm(\cdot;\bk),\mathcal{P}_r(\cdot) \right\rangle_{L^2(\Omega)}\nn\\
&\ \ =\ \frac{1}{\sqrt{2}}\ \frac{\kappa_1+i\kappa_1}{(\kappa_1^2+\kappa_2^2)^{1\over2}}\ \left\langle p_1,\hat{\gamma}_r\left(\ \frac{\kappa}{\delta},\delta s\right)\cdot \mathcal{P}_r \right\rangle_{L^2(\Omega)}\ \pm\ \frac{1}{\sqrt{2}}\ \left\langle p_2,\hat{\gamma}_r\left(\ \frac{\kappa}{\delta},\delta s\right)\cdot \mathcal{P}_r\right\rangle_{L^2(\Omega)}\nn\\
&\qquad\qquad\qquad\ \ +\ \mathcal{O}\Big(|\bkappa|\cdot\|\alpha_0\|_{W^{2,1}(\R^2)}\Big)\nn\\
&\ \ =\ \frac{1}{\sqrt{2}}\ \frac{\kappa_1+i\kappa_2}{(\bkappa_1^2+\bkappa_2^2)^{1\over2}}\ \left\langle \Phi_1,\hat{\gamma}_r\left(\ \frac{\kappa}{\delta},\delta s\right)\cdot \Psi_r \right\rangle_{L^2(\Omega)}\ \pm\ \frac{1}{\sqrt{2}}\ \left\langle \Phi_2,\hat{\gamma}_r\left(\ \frac{\kappa}{\delta},\delta s\right)\cdot \Psi_r\right\rangle_{L^2(\Omega)} \nn\\
&\qquad\qquad\qquad\ \ +\ \mathcal{O}\Big(|\bkappa|\cdot\|\alpha_0\|_{W^{2,1}(\R^2)}\Big)
\label{gamma-innerp}
\end{align}
}}
The inner products in \eqref{gamma-innerp} can be evaluated using the expressions for 
$\hat\gamma_r(\bX,T),\ 1\le r\le 4$ displayed in \eqref{g1}-\eqref{g4} and 
the following mild generalization of Proposition 4.1 of \cite{FW:12} to the case of complex $\zeta=(\zeta_1,\zeta_2)$:
\begin{proposition}\label{mild-gen}
 \begin{align}
\left\langle \Phi_a,\zeta\cdot \nabla\Phi_a\right\rangle &=\ 0,\ \ a=1,2\ ,\nn\\
 2i\ \left\langle \Phi_1,\zeta\cdot\nabla\Phi_2\right\rangle &=\  \overline{2i\ \left\langle \Phi_2,\overline{\zeta}\cdot\nabla\Phi_1\right\rangle}\ =\  -\overline{\lambda_\sharp}\ \times\ (\zeta_1+ i\zeta_2),\nn\\
2i\left\langle \Phi_2,\zeta\cdot\nabla\Phi_1\right\rangle\ &=\  -\lambda_\sharp\ \times\ (\zeta_1- i\zeta_2)
\label{inner-prods}\end{align}
where
\begin{equation}
\lambda_\sharp\ =\ 3\ {\rm area}(\Omega)\ \times\   \sum_{\bm\in\mathcal{S}}\ c(\bm,\mu_\star)^2\ 
 \left(\begin{array}{c}1\\ i\end{array}\right)\cdot \bK^\bm\ \ .
 \label{lambda_sharp}
 \end{equation}
Here, $\{c(\bm;\mu_\star)\}_{\bm\in\mathcal{S}}$ denotes the sequence of $L^2_{\bK,\tau}$ Fourier coefficients of the normalized eigenstate 
 \[\Phi_1\in\ L^2_{\bK_\star,\tau}\ \cap\ \rm{Nullspace}\Big(-\Delta+V_h-\mu_\star\Big)\]
 and $\mathcal S\subset\mathbb Z^2$ is as in \cite{FW:12}.
\end{proposition}
\medskip

\nit The inner products in \eqref{gamma-innerp} can be evaluated, thanks to \eqref{g1}-\eqref{g4}
 and Proposition \ref{mild-gen}. \bigskip

\nit Inner products of the form: $\left\langle \Phi_1\ ,\ \cdot\right\rangle\ :$
\begin{align}
& \left\langle \Phi_1,\widehat{\gamma_1}\cdot \Psi_1 \right\rangle_{L^2(\Omega)}\ =\ 
i\widehat{\D_T\alpha_1},\nn\\ 
&\left\langle \Phi_1,\widehat{\gamma_2}\cdot \Psi_2 \right\rangle_{L^2(\Omega)}\ =\ 0,\nn\\
& \left\langle \Phi_1,\widehat{\gamma_3}\cdot \Psi_3 \right\rangle_{L^2(\Omega)}\ =\ 
 \left\langle \Phi_1,2\widehat{\nabla_\bX\alpha_1}\cdot \nabla_\bx\Phi_1\right\rangle_{L^2(\Omega)}
  \ =\ 0 ,  \nn\\
& \left\langle \Phi_1,\widehat{\gamma_4}\cdot \Psi_4 \right\rangle_{L^2(\Omega)}\ =\     \left\langle \Phi_1,2\widehat{\nabla_\bX\alpha_2}\cdot \nabla_\bx\Phi_2 \right\rangle_{L^2(\Omega)}\ =\ 
 i\overline{\lambda_\sharp}\ \left(\widehat{\D_{X_1}\alpha_2} + i\ \widehat{\D_{X_2}\alpha_2}\right)\ .
 \label{Phi1-ip}
 \end{align}

\nit Inner products of the form: $\left\langle \Phi_2\ ,\ \cdot\right\rangle\ :$
\begin{align}
&\left\langle \Phi_2,\widehat{\gamma_1}\cdot \Psi_1 \right\rangle_{L^2(\Omega)}\ =\ 0\nn\\
&\left\langle \Phi_2,\widehat{\gamma_2}\cdot \Psi_2 \right\rangle_{L^2(\Omega)}\ =\
 i\widehat{\D_T\alpha_2},
\nn\\
&\left\langle \Phi_2,\widehat{\gamma_3}\cdot \Psi_3 \right\rangle_{L^2(\Omega)}\ =\ 
\left\langle \Phi_2,2\widehat{\nabla_\bX\alpha_1}\cdot \nabla_\bx\Phi_1 \right\rangle_{L^2(\Omega)}\ =\ \ i\lambda_\sharp\left(\ \widehat{\D_{\bX_1}\alpha_1}\ -\ i\ \widehat{\D_{\bX_2}\alpha_1}\ \right)\nn\\
 & \left\langle \Phi_2,\widehat{\gamma_4}\cdot \Psi_4 \right\rangle_{L^2(\Omega)}\ =\ 
  \left\langle \Phi_2,2\widehat{\nabla_\bX\alpha_2}\cdot \nabla_\bx\Phi_2 \right\rangle_{L^2(\Omega)}\ =\ 0.\label{Phi2-ip}
\end{align}
In \eqref{Phi1-ip} and \eqref{Phi2-ip} the Fourier transforms are evaluated at $(\frac{\kappa}{\delta},\delta s)$. 
\medskip

Now summing both sides of \eqref{gamma-innerp} over $1\le r\le 4$ we obtain:
\begin{align}
&\sum_{r=1}^4\ \hat{\gamma}_r\left(\ \frac{\bkappa}{\delta},\delta s\right)\  
 \cdot \ 
\left\langle p_\pm(\cdot;\bk),\mathcal{P}_r(\cdot) \right\rangle_{L^2(\Omega)}\nn\\
&=\ \ \frac{i}{\sqrt{2}}\ \frac{\kappa_1+i\kappa_1}{(\kappa_1^2+\kappa_2^2)^{1\over2}}\times
\Big[\ \widehat{\D_T\alpha_1}\ +\   \overline{\lambda_\sharp}\ \left(\widehat{\D_{X_1}\alpha_2} + i\ \widehat{\D_{X_2}\alpha_2}\right)\ \Big]\left(\frac{\bkappa}{\delta},\delta s\right) \nn\\
&\qquad\ \pm\ \frac{i}{\sqrt{2}}\ \Big[\ \widehat{\D_T\alpha_2}\ +\  \lambda_\sharp\ \left(\widehat{\D_{X_1}\alpha_1} - i\ \widehat{\D_{X_2}\alpha_1}\right)\ \Big]\left(\frac{\bkappa}{\delta},\delta s\right)
+ \mathcal{O}\Big(\ |\bkappa|\cdot\|\alpha_0\|_{W^{2,1}(\R^2) }\Big)\nn\\
&=_{\rm def}\  \frac{i}{\sqrt{2}}\ \frac{\kappa_1+i\kappa_1}{(\kappa_1^2+\kappa_2^2)^{1\over2}}\ \cdot\ \widehat{\mathcal{D}_1}\left(\frac{\bkappa}{\delta},\delta s\right)\ \pm \frac{i}{\sqrt{2}}\ \widehat{\mathcal{D}_2}\left(\frac{\bkappa}{\delta},\delta s\right) 
+ \mathcal{O}\Big(\ |\bkappa|\cdot\|\alpha_0\|_{W^{2,1}(\R^2) }\Big)
\label{Dirac-dominates}\end{align}
%
%
\begin{remark}\label{Dirac-Dj}
Note that the Dirac equations \eqref{Dirac-1}-\eqref{Dirac-2} are equivalent to the equations
\begin{equation}
\mathcal{D}_1(\bX,T)\ =\ 0,\qquad\ \mathcal{D}_2(\bX,T)\ =\ 0\ .
\label{D_j=0}
\end{equation}
\end{remark}

Recalling the definition \eqref{ip-re-expressed} of ${\rm Term}_{1,r}$, we find that:
\begin{align}
& \sum_{r=1}^4\  \textrm{Term}_{1,r} \ =\ \delta^{-2}\sum_{r=1}^4\ \hat{\gamma}_r\left(\ \frac{\bkappa}{\delta},\delta s\right)\  
 \cdot \ 
\left\langle p_\pm(\cdot;\bk),\mathcal{P}_r(\cdot) \right\rangle_{L^2(\Omega)}\ 
\chi\Big(|\bk-\bK| < \delta^\tau\ \Big)\nn\\
&\ \ \ \ \ =\  \delta^{-2}\ \chi\left(|\bkappa|<\delta^\tau\right)\ 
 \Big[\ \frac{i}{\sqrt{2}}\ \frac{\kappa_1+i\kappa_1}{(\kappa_1^2+\kappa_2^2)^{1\over2}}\ \cdot\ \widehat{\mathcal{D}_1}\left(\frac{\bkappa}{\delta},\delta s\right)\ \pm \frac{i}{\sqrt{2}}\ \widehat{\mathcal{D}_2}\left(\frac{\bkappa}{\delta},\delta s\right)\ \Big]
 \nn\\
&\ \ \ \  +\  \mathcal{O}\left(\ \delta^{-2}\ \ |\bkappa|\ \chi\left(|\bkappa|<\delta^\tau\right)\cdot\|\alpha_0\|_{W^{2,1}(\R^2) }\right)\ .
\label{inp}\end{align}
By \eqref{eta-pm-split}, \eqref{eta-near-model},  \eqref{ip-re-expressed}, \eqref{int-Erdelta} and \eqref{inp}, for any fixed $t>0$: 
{\footnotesize{
\begin{align}
&\tilde{f}_{I,\pm}(\bk,t)\nn\\
 &= -i\delta^2\ \int_0^t\ e^{-i(\mu_\pm(\bk)-\mu_\star)(t-s)}\nn\\ 
 &\qquad\qquad\qquad\times\ \sum_r\ \left\langle \Phi_\pm(\cdot;\bk) , 
\gamma_r(\delta\cdot,\delta s)\Psi_r(\cdot)\right\rangle_{L^2(\R^2)}\ 
\chi\Big(|\bk-\bK| < \delta^\tau\ \Big)\ ds\nn\\
&= -i\ \int_0^t\ e^{-i(\mu_\pm(\bk)-\mu_\star)(t-s)}\ \nn\\ 
&\qquad\qquad\qquad \chi\left(|\bkappa|<\delta^\tau\right)\ \Big[\ \frac{i}{\sqrt{2}}\ \frac{\kappa_1+i\kappa_1}{(\kappa_1^2+\kappa_2^2)^{1\over2}}\ \cdot\ \widehat{\mathcal{D}_1}\left(\frac{\bkappa}{\delta},\delta s\right)\pm \frac{i}{\sqrt{2}} \widehat{\mathcal{D}_2}\left(\frac{\bkappa}{\delta},\delta s\right)\ \Big]\Big|_{\bkappa=\bk-\bK} ds
\ \nn\\
 &\ \ +\ \mathcal{O}\Big(\ |t|\ 
\ |\bk-\bK|\ \ \chi\left( |\bk-\bK|<\delta^\tau\right)\cdot\|\alpha_0\|_{W^{2,1}(\R^2)} \Big)\nn\\
&\ \ \ \   +\ \mathcal{O}\Big(\ |t|\ \ \chi\left( |\bk-\bK|<\delta^\tau\right)\  \delta^{A}\cdot\|\alpha_0\|_{W^{A+1,1}}\ \Big)
\label{eta-near123}
\end{align} 
}}
Recall that
\begin{align}
\| f_{I,D}(t) \|_{L^2(\R^2)}^2\ &=\ \sum_{b=\pm}\ 
\int_{\mathcal{B}_h}\ \chi\Big(|\bk-\bK|<\delta^\tau\Big)\ |\tilde f_b(\bk,t)|^2\ d\bk\nn\\
&\ =\ \sum_{b=\pm}\ 
\int_{\mathcal{B}_h}\  |\tilde f_{I,b}(\bk,t)|^2\ d\bk
\label{eta-near-A}\end{align}

\begin{remark}\label{motivate-dirac}
 In this remark we 
 assess the contribution of the first term on the right hand side of the upper bound  \eqref{eta-near123} to $\| f_{I,D}(t) \|_{L^2(\R^2)}^2$ for times  $ t\approx \rho\ \delta^{-1-\eps_0}$. 
and argue that the Dirac equations \eqref{Dirac-1}-\eqref{Dirac-2} are necessary
 for $f^\delta(\bx,t)$ to be small on a time scale of the order $\delta^{-1}$.
 
 This contribution to the right hand side of \eqref{eta-near-A} is bounded by a constant multiple of
\begin{align}
&|t|^2\  \int_{\mathcal{B}_h}\ \chi\Big(|\bK-\bk| <\delta^\tau\Big)\ \ d\bk\ \approx\ \left(\ \rho\ \delta^{-1-\eps_0}\ \right)^2\ \delta^{2\tau} \approx \left(\delta^{-(1-\tau +\eps_0)}\right)^2
\nn \end{align}
Recall that $0<\tau<1$ and therefore this contribution diverges as $\delta\downarrow0$. Indeed this is the case even for $t\approx \delta^{-1},\ (\eps_0=0)$. We conclude that for $f$ to be small in $L^2(\R^2)$ for $\delta\downarrow0$ we must have
\begin{align}
&\chi\left(|\bkappa|<\delta^{\tau-1}\right)\ 
\Big[\ \widehat{\D_T\alpha_1}\ +\  \ \overline{\lambda_\sharp}\ \left(\widehat{\D_{X_1}\alpha_2} + i\ \widehat{\D_{X_2}\alpha_2}\right)\ \Big]\left(\frac{\bkappa}{\delta},\delta t\right)\ =\ 0\nn\\
&\chi\left(|\bkappa|<\delta^{\tau-1}\right)\ \Big[\ \widehat{\D_T\alpha_2}\ +\ \lambda_\sharp\ \left(\widehat{\D_{X_1}\alpha_1} - i\ \widehat{\D_{X_2}\alpha_1}\right)\ \Big]\left(\frac{\bkappa}{\delta},\delta t\right)\ =\ 0 .
\label{FT-Dirac}\end{align}
Indeed, these are implied by the Dirac equations \eqref{D_j=0} or equivalently \eqref{Dirac-1}-\eqref{Dirac-2}.
\end{remark}
\bigskip

Since the Dirac equations are assumed to hold, $\tilde f_{I,\pm}(\bk,t)$ is controlled  by the latter two terms in 
\eqref{eta-near123}. Their contributions to  $\|f_{I,D}(t)\|_{L^2(\R^2)}^2$ are bounded, for $0\le t\le \rho\ \delta^{-2+\eps_1}$, as follows. Fix $\eps_1>0$. The second term in \eqref{eta-near123} gives a contribution to $\|f_{I,D}(t)\|_{L^2(\R^2)}^2$  which is  bounded by:
\begin{align}
&|t|^2\  \int_{\mathcal{B}_h} |\bk-\bK|^2\ \chi\Big(|\bk-\bK|<\delta^\tau\Big)\ \ d\bk\nn\\
 &\approx\ |t|^2\ \delta^{4\tau}\ \le\ \delta^{2\eps_1-4(1-\tau)} \le \delta^{\eps_1} 
\end{align}
 by taking $\tau$ chosen sufficiently close to $1$. 
The third term in \eqref{eta-near123}, for $A$ sufficiently large, clearly
  gives a contribution to $\|f_{I,D}(t)\|_{L^2(\R^2)}^2$  which is  $\mathcal{O}(\delta^{\eps_1})$ for $0\le t\le \rho\ \delta^{-2+\eps_1},\ \eps_1>0$ as $\delta\downarrow0$.
  \medskip
  
  In conclusion, for any fixed $\rho>0$ and $\eps_1>0$ , we have
  \begin{equation}
  \sup_{0 \le t \le  \rho\delta^{-2+\eps_1}}\ \| f_{I,D}(t) \|_{L^2(\R^2)}\ =\ \mathcal{O}(\delta^{\eps_1\over2})\ .
  \label{fI-pm-est}
  \end{equation}
  This proves the bound \eqref{f1-est}.

\subsection{Estimation of $\|f_{II,D}(t)\|_{L^2(\R^2)}$}\medskip

\qquad{\ }\medskip

In this section we prove that for any fixed $\rho>0$ and $\eps_1>0$
\begin{align}
  & \sup_{0\le t \le \rho\delta^{-2+\eps_1}}\|f_{II,D}(\cdot,t)\|_{L^2(\R^2)}\ =\ \mathcal{O}(\delta^{\eps_1}),\ \ \textrm{as}\ \delta\downarrow0
   \label{f2-est}\end{align}

Since
\[
f_{II,D}(\bx,t)\ =\ \sum_{b\in\{+,-\}}\ \int_{\mathcal{B}_h}\ 
\chi\Big(|\bk-\bK|\ge\delta^\tau\Big)\ \tilde f_b(\bk,t)\ \Phi_b(\bx;\bk)\ d\bk\ ,
\]
 we have 
 \begin{equation}
 \|f_{II,D}(\cdot,t)\|_{L^2(\R^2)}^2\ =\ \sum_{b\in\{+,-\}}\ \int_{\brill_h}\ 
\chi\Big(|\bk-\bK|\ge\delta^\tau\Big)\  |\tilde f_b(\bk,t)|^2\ d\bk
\label{fII-norm} \end{equation}
where we recall again, by \eqref{teta-far}, that 
\begin{equation}
\tilde f_\pm(\bk,t) = -i\delta^2\ \int_0^t\ ds\ e^{-i(\mu_\pm(\bk)-\mu_\star)(t-s)}\ 
 \sum_{r=1}^4\ \left\langle \Phi_\pm(\cdot;\bk) , \gamma_r(\delta\cdot,\delta s)\Psi_r(\cdot)\right\rangle_{L^2(\R^2)}\ .
\label{eta-near-model-1}
 \end{equation}

\nit By Proposition \ref{p-sum-ip} with the choices: $b=+$ or $b=-$\ , $\Gamma=\gamma_r(\delta\bx,\delta s)$ and $\mathcal{P}=\mathcal{P}_r$ we have 
\begin{align}
& \left\langle \Phi_\pm(\cdot;\bk) , 
\gamma_r(\delta\cdot,\delta s)\Psi_r(\cdot)\right\rangle_{L^2(\R^2)}\nn\\
&=\ \int_\Omega\ \overline{p_\pm(\by;\bk)}\ \Big[\ \delta^{-2}\ \sum_{\bm\in\Z^2}\ e^{i\bm\cdot\by}\ \hat{\gamma}_r\left(\frac{m_1\bk_1+m_2\bk_2+(\bk-\bK)}{\delta},\delta s\right)\ \Big]\ 
  \mathcal{P}_r(\by)\ d\by \label{ip-100}
  \end{align}
  The next lemma implies that all terms in the infinite sum within \eqref{ip-100} involve $\hat\gamma_r$  evaluated 
  at a large argument.    \medskip
  
  \begin{lemma}\label{dist} 
  Let $\bk\in\mathcal{B}_h$ and assume
$|\bk-\bK|\ge\delta^\tau$. 
   Then, there exists a constant, $c>0$, such that  for all $\bm=(m_1,m_2)\in\Z^2$
  \[ \left|\ m_1\bk_1+m_2\bk_2 + (\bk-\bK)\ \right|\ \ge\ c\ \delta^\tau\ \Big( 1 + |\bm| \Big). \]
  \end{lemma}
{\it   Proof of Lemma \ref{dist}:}\ 
  $|m_1\bk_1+m_2\bk_2 + (\bk-\bK)|$ is equal to the distance from the point $\bK-m_1\bk_1-m_2\bk_2$ to $\bk$. Simple geometry concludes the proof.\bigskip
  
  By Proposition \ref{gammahat-est} and Lemma \ref{dist} , for $\bk\in\mathcal{B}_h$ and 
  $|\bk-\bK|\ge\delta^\tau$ we have 
  \begin{align}
&\left|\  \delta^{-2}\ \sum_{\bm\in\Z^2}\ e^{i \bm \cdot\by}\ \hat{\gamma}_r\left(\frac{m_1\bk_1+m_2\bk_2+(\bk-\bK)}{\delta},\delta s\right)\ \right|\nn\\
&\ \ \ \le\ C\ \sum_{\bm\in\Z^2}\frac{1}{(1+|\bm|)^A}\ \delta^{-2}\ (\delta^{1-\tau})^A\  \|\alpha_0\|_{W^{A+1,1}(\R^2)}\ .\nn
\end{align}
From \eqref{eta-near-model-1} and \eqref{ip-100}, we have that for $\bk\in\mathcal{B}_h$ and 
  $|\bk-\bK|\ge\delta^\tau$ by taking $A>2$:
  \begin{align}
\left|  \tilde f_\pm(\bk,t) \right|\ \le\ C\ |t|\ (\delta^{1-\tau})^A\ \|\alpha_0\|_{W^{A+1,1}(\R^2)}\ .
  \end{align}
  By \eqref{fII-norm}, for $0\le t\le\delta^{-2+\eps_1}$, by taking $A\ge A_0\ge2$ sufficiently large, we have the bound $ \|f_{II,\pm}(\cdot,t)\|_{L^2(\R^2)}=\mathcal{O}(\delta^{10})$, from which \eqref{f2-est} follows. Recalling  \eqref{fI-pm-est}, we obtain
  \begin{equation}
  \sup_{0\le t\le\delta^{-2+\eps_1}}\ \|f_D(t)\|_{L^2(\R^2)}\ =\ \mathcal{O}(\delta^{\frac{\eps_1}{2}})\ .
  \label{fD-est}
  \end{equation}

\subsection{Estimation of $\|f_{D^c}(\cdot,t)\|_{H^s(\R^2)}$}\label{eta-far-estimates}
\bigskip

{\qquad}\medskip

 Again, we fix $\rho>0$ and $\eps_1>0$ and assume $0\le t\le \rho\delta^{-2+\eps_1}$. Recall from \eqref{eta-far} that  
 
 \begin{align}
 f_{D^c}(\bx,t)\ &=\  \sum_{ b\notin\{+,-\}}\ \int_{\mathcal{B}_h}\ 
 \ \tilde f_b(\bk,t)\ \Phi_b(\bx;\bk)\ d\bk\ ,
\label{eta-far1}\\ \tilde f_b(\bk,t) &= -i\delta^2\ \int_0^t\ ds\ e^{-i(\mu_b(\bk)-\mu_\star)(t-s)}\ 
 \left\langle \Phi_b(\cdot;\bk) , \sum_{r=1}^4\gamma_r(\delta\cdot,\delta s)\Psi_r(\cdot)\right\rangle_{L^2(\R^2)} \ .
 \nn\\
 & \label{teta-far1}
\end{align}

To prove that 
\[ \sup_{0\le t\le \rho \delta^{-2+\eps_1}}\ \|f_{D^c}(\cdot,t)\|_{H^s(\R^2)}\  =
\mathcal{O}\left(\delta^{\frac{\eps_1}{2}}\right),\ \  \delta\downarrow0,\]
 we shall decompose $f_{D^c}$ into
its $\bk$- components near and away from $\bK$. For $\bk$ near $\bK$, Property (P2), \eqref{mupm-lowerbd}, implies that the complex exponential in \eqref{teta-far1} is oscillatory
 and so integration by parts gains us smallness via additional powers of  $\delta$. For $\bk\in\mathcal{B}_h$  but away from $\bK$ we use that the Fourier transform of  $\gamma_r(\delta\bx,\delta s)\Psi_r(\bx)$ is mainly supported away from $\bK$. Finally, smoothness  of $\gamma_r(\delta\bx,\delta s)\Psi_r(\bx)$ is used to ensure sufficient decay as $b\to\infty$, of its $\Phi_b(\cdot;\bk)$- components, which must be summed in order to control $H^s$ norms. 
 
To implement the above strategy we decompose $f_{D^c}$ as
\begin{align} 
f_{\notDr}(\bx,t)\ &=\  \sum_{ b\notin\{+,-\}}\ \int_{\mathcal{B}_h}\ 
\chi\Big(|\bk-\bK|<\kappa_1\Big)\ \tilde f_b(\bk,t)\ \Phi_b(\bx;\bk)\ d\bk\nn\\
 &\ \ \ \  +\
  \sum_{ b\notin\{+,-\}}\ \int_{\mathcal{B}_h}\ 
 \chi\Big(|\bk-\bK|\ge\kappa_1\Big)\  \tilde f_b(\bk,t)\ \Phi_b(\bx;\bk)\ d\bk\nn
 \\
 &=\ f_{I,\notDr}(\bx,t)\ +\ f_{II,\notDr}(\bx,t)
\label{eta-far2}
\end{align}
 
 By property (P2), \eqref{mupm-lowerbd}, we have
\begin{equation}
 |\bk-\bK|<\kappa_1\ \ \ \implies\ \ \left|\mu_b(\bk)-\mu_\star\right|\ge C_1\
  \textrm{for}\  b\ne\pm .\label{P2implies}
 \end{equation}
 It is natural to exploit the oscillation coming from the complex exponential in \eqref{teta-far1}. 
 Fix $\bk\in\mathcal{B}_h$
  and such that  $|\bk-\bK|<\kappa_1$. By \eqref{P2implies}, we may integrate by parts once and obtain:
 {\small{
    \begin{align}
  \tilde f_b(\bk,t)\ &=\ \delta^2\ \frac{e^{-i(\mu_b(\bk)-\mu_\star)t}-1}{\mu_b(\bk)-\mu_\star}\ \left\langle \Phi_b(\cdot;\bk), \sum_{r=1}^4\gamma_r(\delta\cdot,0)\ \Psi_r(\cdot)\right\rangle_{L^2(\R^2)}\nn\\
  &\ \ +\ i\ \delta^3\ \int_0^t\ ds\ 
  \frac{e^{-i(\mu_b(\bk)-\mu_\star)(t-s)}-1}{\mu_b(\bk)-\mu_\star}\ 
  \left\langle \Phi_b(\cdot;\bk), \sum_{r=1}^4\D_T\gamma_r(\delta\cdot,\delta s)\ \Psi_r(\cdot)\right\rangle_{L^2(\R^2)}
 \end{align}
 }}
%
 Therefore
 \begin{align}
  |\tilde f_b(\bk,t)|\ &\lesssim\ \delta^{2}\ \frac{1}{|\mu_b(\bk)-\mu_\star|}\ 
\left| \left\langle \Phi_b(\cdot;\bk),\sum_{r=1}^4\gamma_r(\delta\cdot,0)\ \Psi_r(\cdot)\right\rangle_{L^2(\R^2)}\right|\nn\\
&+\  \delta^{3}\  \int_0^t\ ds\ \frac{1}{|\mu_b(\bk)-\mu_\star|}\ 
  \left|\left\langle \Phi_b(\cdot;\bk), \sum_{r=1}^4\D_T\gamma_r(\delta\cdot,\delta s)\ \Psi_r(\cdot)\right\rangle_{L^2(\R^2)}\right|
\label{fb-smallindelta} \end{align}
%
 For  $|\bk-\bK|<\kappa_1$  and $b\ne\pm$, we have by \eqref{P2implies}

\begin{align}
& |\tilde f_b(\bk,t)|\ \lesssim \delta^2\ 
\left| \left\langle \Phi_b(\cdot;\bk), \sum_{r=1}^4\gamma_r(\delta\cdot,0)\ \Psi_r(\cdot)\right\rangle_{L^2(\R^2)}\right|\nn\\
&\ \ \ \ \ +\ \delta^3\  |t|\ 
 \max_{0\le s\le \rho\delta^{-2+\eps_0}}\ \left|\left\langle \Phi_b(\cdot;\bk), \sum_{r=1}^4\D_T\gamma_r(\delta\cdot,\delta s)\ \Psi_r(\cdot)\right\rangle_{L^2(\R^2)}\right|
 \label{tfb-thing}\end{align} 
Therefore, for $|\bk-\bK|<\kappa_1$  and $b\ne\pm$
\begin{align}
&\sup_{0\le t\le \rho\delta^{-2+\eps_0}}\ |\tilde f_b(\bk,t)|^2\nn\\
&\qquad \lesssim\  \delta^4\ \left| \left\langle \Phi_b(\cdot;\bk), \sum_{r=1}^4\gamma_r(\delta\cdot,0)\ \Psi_r(\cdot)\right\rangle_{L^2(\R^2)}\right|^2\nn\\
&\qquad\qquad +\  \delta^{2+2\eps_0}\  \max_{0\le s\le \rho\delta^{-2+\eps_0}}\ \left|\left\langle \Phi_b(\cdot;\bk), \sum_{r=1}^4\D_T\gamma_r(\delta\cdot,\delta s)\ \Psi_r(\cdot)\right\rangle_{L^2(\R^2)}\right|^2.
\label{fb-1}\end{align}

 To obtain a  bound on  $\| f_{I,D^c}(\cdot, t)\|_{H^s(\R^2)}$ for any $s>0$, we proceed as follows. Recall
 \begin{equation}
 \| f_{I,D^c}(\cdot, t)\|^2_{H^s(\R^2)}\ \sim\ 
 \sum_{b\notin\{+,-\}}\ \left(1+|b|\right)^s\int_{\Big\{\bk\in\brill_h: |\bk-\bK|<\kappa_1\Big\}} |\tilde f_b(\bk,t)|^2\ d\bk
\label{fIDc} \end{equation}
 {\it  By the sum over $b\notin\{+,-\}$ we mean the sum over  all $b\ge1$ such that 
 $b\notin\{b_1,b_1+1\}$, where
  $\mu_{b_1}(\bk)\equiv\mu_-(\bk)$ and 
 $\mu_{b_1+1}(\bk)\equiv\mu_+(\bk)$; see Definition \ref{Diracpt-def}.}
 \medskip
 
\nit Thus we'll require decay of  $|\tilde f_b(\bk,t)|$ for $b$ large. This decay is obtained from the inner products in \eqref{fb-1}. Observe, for some sufficiently large postive constant, $\tilde C$ and any $M\ge0$ and $j\ge0$:
  {\small{
  \begin{align}
& \left\langle \Phi_b(\cdot;\bk), \D_T^j\gamma_r(\delta\cdot,\delta s)\ \Psi_r(\cdot)\right\rangle_{L^2(\R^2)}\nn\\
& =\ \frac{1}{(\tilde C+\mu_b(\bk))^M}\left\langle (\tilde C+H)^M\Phi_b(\cdot;\bk), \D_T^j\gamma_r(\delta\cdot,\delta s)\ \Psi_r(\cdot)\right\rangle_{L^2(\R^2)}\nn\\
 &=\ \frac{1}{(\tilde C+\mu_b(\bk))^M}\left\langle \Phi_b(\cdot;\bk), (\tilde C+H)^M\left(\D_T^j\gamma_r(\delta\cdot,\delta s)\ \Psi_r(\cdot) \right) \right\rangle_{L^2(\R^2)}\nn
 \end{align}
 Therefore, thanks to \eqref{gammaHs}: 
 \begin{align}
 \left|\left\langle \Phi_b(\cdot;\bk), \D_T^j\gamma_r(\delta\cdot,\delta s)\ \Psi_r(\cdot)\right\rangle_{L^2(\R^2)}\right|\ &\lesssim\ (1+|b|)^{-M}\ \left\| \gamma_r(\delta\cdot,\delta s)\ \Psi_r(\cdot)\ \right\|_{H^{2M+j}(\R^2)}\nn\\
 & \lesssim\ (1+|b|)^{-M}\ \delta^{-1}\  \|\alpha_0\|_{H^{2M+j}(\R^2)}
 \label{b-est}
  \end{align}
  }}
  for $j\ge0$.
  Using \eqref{b-est} in \eqref{fb-1} we obtain 
  \begin{align}
&\sup_{0\le t\le \rho\delta^{-2+\eps_0}}\ |\tilde f_b(\bk,t)|^2\ \lesssim\ 
 (1+|b|)^{-2M}\   \|\alpha_0\|_{H^{2M+1}(\R^2)}^2\ \delta^{2\eps_0}\ 
\label{fb-2}\end{align}
  Substituting into \eqref{fIDc},  we get 
  \begin{align}
  \| f_{I,D^c}(\cdot, t)\|_{H^s(\R^2)}\ &\sim\ 
\Big( \sum_{b\notin\{+,-\}}\ \left(1+|b|\right)^{s-2M}\Big)^{1\over2}\ \|\alpha_0\|_{H^{2M+1}(\R^2)}\ \delta^{\eps_0}\nn\\
 &\lesssim\  \|\alpha_0\|_{H^{2M+1}(\R^2)}\ \delta^{\eps_0}
\label{fIDc-A} \end{align}
provided $2M>s+1$.
\bigskip

It remains to estimate the $H^s$ norm of
\begin{equation}
 f_{II,D^c}(\cdot;t)\ =\ 
 \sum_{ b\notin\{+,-\}}\ \int_{\mathcal{B}_h}\ 
 \chi\Big(\ |\bk-\bK|\ge\kappa_1\Big)\ \tilde f_b(\bk,t)\ \Phi_b(\bx;\bk)\ d\bk,
 \label{fII-far1}
 \end{equation}
 where we recall
 \begin{equation}
 \| f_{II,D^c}(\cdot;t)\|_{H^s}^2\ \sim\ \sum_{b\not\in\{+,-\}}\ (1+|b|)^s\ \int_{\mathcal{B}_h}\ 
 \chi\Big(\ |\bk-\bK|\ge\kappa_1\Big)\  |\tilde f_b(\bk,t)|^2\ d\bk\ .
 \label{fIIDc}\end{equation}
 Note that from \eqref{teta-far1}
 {\small{
 \begin{align}
& \int_{\mathcal{B}_h}\ 
 \chi\Big(\ |\bk-\bK|\ge\kappa_1\Big)\ |\tilde f_b(\bk,t)|^2\ d\bk\nn\\ 
& \le \delta^4 |t|^2\ \max_{0\le s\le t}\ \int_{\mathcal{B}_h}\ 
 \chi\Big(\ |\bk-\bK|\ge\kappa_1\Big)\
\  \left| \sum_{r=1}^4 \left\langle \Phi_b(\cdot;\bk) , 
 \gamma_r(\delta\cdot,\delta s)\ \Psi_r(\cdot)\right\rangle_{L^2(\R^2)} \right|^2\ d\bk
 \label{bth-term}\end{align}
 }}
 A crude bound on \eqref{bth-term}, valid for $0\le t\le \delta^{-2+\eps_0}$ uses the approach taken to obtain \eqref{b-est}. This gives the bound: $\delta^4\ (\delta^{-2+\eps_0})^2\ \delta^{-2}\ \|\alpha_0\|^2_{W^{2,1}(\R^2)}$, which becomes unbounded as $\delta\downarrow0$. Therefore,  a sharper estimate is required.
 
 Now, for $\bk\in\brill,\ |\bK-\bk|\ge\kappa_1$ and $b\ne\pm$,  it may be that $|\mu_b(\bk)-\mu_\star|$ is small.
 Hence  the integral in \eqref{teta-far1} cannot be controlled as an oscillatory integral, via integration by parts
  with respect to time. We therefore obtain the decay of  $ f_{II,D^c}$ using the rapid decay of 
 the Fourier transform of $\gamma_r$.\medskip
 
 The inner product can be rewritten and estimated as follows. Choose a positive constant $\tilde C$ such that $\tilde C I + H$ is strictly  positive. We have, for any integer $M\ge0$,

 {\footnotesize{
 \begin{align}
&  \left\langle \Phi_b(\cdot;\bk) , 
 \gamma_r(\delta\cdot,\delta s)\cdot \Psi_r(\cdot)\right\rangle_{L^2(\R^2)} \nn\\
 &=\  \frac{1}{(\tilde C+\mu_b(\bk))^M}\ \left\langle (\tilde C+H)^M\Phi_b(\cdot;\bk), \gamma_r(\delta\cdot,\delta s)\ \Psi_r(\cdot)\right\rangle_{L^2(\R^2)}\nn\\
 &=\ \frac{1}{(\tilde C+\mu_b(\bk))^M}\ \left\langle \Phi_b(\cdot;\bk), (\tilde C+H)^M\left[\ \gamma_r(\delta\cdot,\delta s)\ \Psi_r(\cdot)\ \right]\ \right\rangle_{L^2(\R^2)}\nn\\
 &=\ \frac{1}{(\tilde C+\mu_b(\bk))^M}\ \widetilde\sum_{a,b_1,b_2}\ \delta^{|b_1|}\  \left\langle \Phi_b(\cdot;\bk),\Pi_{\nu=1}^{\nu_{max}}\D^{a_\nu}_\bx V(\cdot)\ \D_\bX^{b_1} \gamma_r(\delta\cdot,\delta s)\ \D_\bx^{b_2}\Psi_r(\cdot)\ \right\rangle_{L^2(\R^2)},
 \label{ibp-ip}
\end{align}
 }}
 where $\widetilde\sum$ denotes a finite sum  over terms of the above form with $a_1,\dots,a_{\nu_{max}},b_1,b_2$ in $\Z^2_+$ and $ 2\nu_{max}+\sum_{\nu=1}^{\nu_{max}}|a_\nu|+|b_1|+|b_2| \le 2M$. Each inner product  of this sum can be re-expressed, via Poisson summation, using Proposition \ref{p-sum-ip}. With the choices
 \begin{align*}
  \Gamma(\bX,T)&=\D_X^{b_1}\gamma_r(\delta\bx,\delta s),\nn\\
   \ \Psi(\bx)&=\Pi_{\nu=1}^{\nu_{max}}\D^{a_\nu}_\bx V(\bx)\ \D_\bx^{b_2}\Psi_r(\bx)=e^{i\bK\cdot\bx}\ \Pi_{\nu=1}^{\nu_{max}}\D^{a_\nu}_\bx V(\bx)\ (\D_\bx+i\bK)^{b_1}\mathcal{P}_r(\bx)\in L^2_\bK 
   \end{align*}
 we apply Proposition \ref{p-sum-ip} and obtain
 {\small{
 \begin{align}
& \left\langle \Phi_b(\cdot;\bk),\Pi_{\nu=1}^{\nu_{max}}\D^{a_\nu}_\bx V(\cdot)\ \D_\bX^{b_1} \gamma_r(\delta\cdot,\delta s)\ \D_\bx^{b_2}\Psi_r(\cdot)\ \right\rangle_{L^2(\R^2)}\nn\\
 &= \int_\Omega\ \overline{p_b(\by;\bk)}\cdot\ \Big[\ \delta^{-2}\ \sum_{\bm\in\Z^2}\ e^{i\bm\cdot\by}\ \widehat{\D^{b_1}_\bX\gamma_r}\left(\frac{m_1\bk_1+m_2\bk_2+(\bk-\bK)}{\delta},\delta s\right)\ \Big]\nn\\  
&\qquad\qquad\qquad\qquad\qquad\times \Pi_{\nu=1}^{\nu_{max}}\D_\by^{a_\nu}V(\by)(\D_\by+i\bK)^{b_2}  \mathcal{P}_r(\by)\ d\by\ .
\label{poisson-ip-gen} 
\end{align}
}}
Note that there is a constant $c>0$, depending only on $\bkappa_1$, such that for all $\bk\in\brill_h$ satisfying $|\bk-\bK|\ge\kappa_1$,  we have  $|m_1\bk_1+m_2\bk_2+(\bk-\bK)|\ge c(1+|\bm|)$. It follows from  Proposition \ref{gammahat-est} applied to \eqref{poisson-ip-gen} that for all $b_1$ and all $A$ such that 
$A\ge M$  and $A>2$:
{\small{
\begin{align}
&\left|\ \left\langle \Phi_b(\cdot;\bk),\Pi_{\nu=1}^{\nu_{max}}\D^{a_\nu}_\bx V(\cdot)\ \D_\bX^{b_1} \gamma_r(\delta\cdot,\delta s)\ \D_\bx^{b_2}\Psi_r(\cdot)\ \right\rangle_{L^2(\R^2)}\ \right|\nn\\
&\qquad \lesssim\  
 \delta^A\ \|\alpha_0\|_{W^{A+1,1}(\R^2)}\ \sum_{\bm\in\Z^2}\ \frac{1}{(1+|\bm|)^A}\ \lesssim\  \delta^A\ \|\alpha_0\|_{W^{A+1,1}(\R^2)}\ .\label{bound1}\end{align}

}}
Recall that $\mu_b(\bk)\approx b$, for $b$ large, uniformly in $\bk\in\brill_h$. Using \eqref{ibp-ip} and \eqref{bound1}
  in \eqref{fIIDc} and \eqref{bth-term} we obtain 
\begin{align}
 \| f_{II,D^c}(\cdot;t)\|_{H^s}^2\ \lesssim\ 
\delta^4\ |t|^2\ \delta^{2A}\ \|\alpha_0\|_{W^{A+1,1}(\R^2)}^2\ \sum_{b\ge1}\ (1+b)^{s-2M}
\nn\end{align}
Choosing $A$ and $M$ sufficiently large, we obtain 
\begin{equation}
\sup_{0\le t\le \delta^{-2+\eps_0}}\ \| f_{II,D^c}(\cdot;t)\|_{H^s}\lesssim\delta^{10}\ \|\alpha_0\|_{W^{A+1,1}(\R^2)}\ \ .\label{f2Dc-est}
\end{equation}
Together, estimates \eqref{fIDc-A} and \eqref{f2Dc-est} imply
\begin{equation}
\sup_{0\le t\le \delta^{-2+\eps_0}}\ \| f_{D^c}(\cdot;t)\|_{H^s}\lesssim\delta^{\eps_0}\ \|\alpha_0\|_{W^{A+1,1}(\R^2)}\ \  .
\label{fDc-est}
\end{equation}

Finally, \eqref{fD-est}, \eqref{fDc-est} and \eqref{cor-Hs} imply, for any $\eps_0>0, \rho>0$, that
\begin{equation}
\sup_{0\le t\le \rho\delta^{-2+\eps_0}}\ \|f^\delta(t)\|_{H^s(\R^2)}\ \lesssim\ \delta^{\frac{\eps_0}{2}}\ \|\alpha_0\|_{W^{A+1,1}(\R^2)}
\label{fdelta-est}
\end{equation}
 This completes the proof of Proposition \ref{average-fast} and therewith Theorem \ref{effective-Dirac}.
 \appendix
 
 \section{Lipschitz-continuity of eigenvalues self-adjoint $2^{nd}-$ order elliptic operators and an application to Floquet-Bloch eigenvalues}\label{EigLip}
 \bigskip
 
 \begin{theorem}\label{lip-continuity}
 Let $T$ and $\tT$ denote operators with the following properies:
 \begin{enumerate}
 \item $T$ and $\tT$  non-negative and self-adjoint operators on $L^2$.
 \item $T$ and $\tT$ are bounded maps from $H^2$ to $L^2$. 
\item $T$ and $\tT$ have discrete spectrum given, respectively, by the sequences of eigenvalues:
 \begin{align}
 &{\rm spec(T)}:\ \ \lambda_1(T)\le\lambda_2(T)\le\cdots\nn\\
& {\rm spec(\tT)}:\ \ \lambda_1(\tT)\le\lambda_2(\tT)\le\cdots\nn\\
 \end{align}
 \item There is a positive constant, $a^\sharp$, such that $T$ satisfies the elliptic estimate:
 \begin{equation}
 \|T\varphi\|_{L^2}+\|\varphi\|_{L^2}\ge a^\sharp\|\varphi\|_{H^2}
 \label{Telliptic}\end{equation}
 for all $\varphi\in H^2$.
 Assume that 
 \begin{equation}
 \|T-\tT\|_{H^2\to L^2}\le \frac{1}{4}a^\sharp
 \label{T-tT}\end{equation}
 \end{enumerate}
 Then, for $k=1,2,\dots$ we have the following Lipschitz estimate
 \begin{equation}
 \left|\lambda_k(T)-\lambda_k(\tT)\right|\ \le\ \frac{2}{a^\sharp}\left(\ 2\lambda_k(T)+1\ \right)\cdot \|\tT-T\|_{H^2\to L^2}
 \label{Lip-est-TmtT}
 \end{equation}
 \end{theorem}

\begin{remark}\label{gen-lip}
 Theorem \ref{lip-continuity} generalizes in a straightforward manner to higher order self-adjoint elliptic operators, defined on $H^s(\R^d)$.
\end{remark}
\medskip
 
 A consequence of Theorem \ref{lip-continuity} is the Lipschitz continuity of the Floquet-Bloch eigenvalues of $H=-\Delta+V(\bx)$, where $V$ is periodic, real and bounded.\medskip
 
 \begin{corollary}[Proposition \ref{ktomuk}]\label{lip-continuity-fb-eigs}
 The Floquet-Bloch eigenvalue maps $\bk\to\mu_b(\bk),\ b\ge1$ are Lipschitz continuous functions of $\bk\in\brill$.
 \end{corollary}
 \medskip
 
 \noindent{\it Proof of Corollary \ref{lip-continuity-fb-eigs}:} 
 Define 
 \[ T=H(\bk_1)=-(\nabla+i\bk_1)^2+V(\bx)\ \ {\rm and}\ \ \tT=H(\bk_2)=-(\nabla+i\bk_2)^2+V(\bx)\]
 and note that $\|T-\tT\|_{H^2\to L^2}=\| H(\bk_1)-H(\bk_2) \|_{H^2\to L^2}\ \le C\ |\bk_1-\bk_2|$ for some $C>0$.
 Let $\mu_b(\bk_j)=\mu_b(H(\bk_j))$ denote the $b^{th}$ eigenvalue of $H(\bk_j)$. Applying 
  Theorem \ref{lip-continuity}, we have
  \begin{equation}
  \left| \mu_b(\bk_1)-\mu_b(\bk_2) \right|\ \le\ C\ (\ |\mu_b(\bk)|+1\ )\ |\bk_1-\bk_2|
  \label{lip-est}
  \end{equation}
  whenever $|\bk_1-\bk_2|$ is less than a small enough positive constant. 
 This completes the proof of Corollary \ref{lip-continuity-fb-eigs}.\bigskip
 
 \noindent{\it Proof of Theorem \ref{lip-continuity}:}\ Note first that 
 \begin{align}
 \|\tT\varphi\|_{L^2}+\|\varphi\|_{L^2} &\ge \left[\ \|T\varphi\|_{L^2}+\|\varphi\|_{L^2}\ \right]
 \ -\ \|T-\tT\|_{H^2\to L^2}\|\varphi\|_{H^2}\nn\\
 &\ge a^\sharp\|\varphi\|_{H^2} \ -\ \|T-\tT\|_{H^2\to L^2}\|\varphi\|_{H^2}\ge \frac{3}{4}a^\sharp\|\varphi\|_{H^2}\ \  .
 \end{align}
 Therefore, we have the following three estimates:
 \begin{align}
 \|T\varphi\|_{L^2}+\|\varphi\|_{L^2}\ge \frac{3}{4}a^\sharp\|\varphi\|_{H^2}\label{TTa}\\
  \|\tT\varphi\|_{L^2}+\|\varphi\|_{L^2}\ge \frac{3}{4}a^\sharp\|\varphi\|_{H^2}\label{TTb}\\
   \|T-\tT\|_{H^2\to L^2}\le \frac{1}{4}a^\sharp\label{TTc}
   \end{align}
   These conditions, which are symmetric in $T$ and $\tT$, will be used below.
   \medskip
   
   Recall now the {\it min-max} characterization of the $k^{th}$ eigenvalue, $\lambda_k(A)$, of a self-adjoint operator, $A:H^2\to L^2$ \cite{Courant-Hilbert-I}:
   \begin{equation}
   \lambda_k(A)\ =\ \min_{S\subset H^2,\ dim(S)=k}\ \  \max_{v\in S\setminus\{0\}}\frac{\left\langle Av,v\right\rangle}{\|v\|_{L^2}^2}
  \label{min-max} \end{equation}
   
   \begin{proposition}\label{H2L2constraint}
   Let $A$ denote a self-adjoint operator which maps $H^2$ to $L^2$ satisfying the ellipticity estimate
   \eqref{TTa}. 
   \begin{equation}
   \lambda_k(A)\ =\ 
   \min_{\substack{S\subset H^2,\ dim(S)=k\\ \|\psi\|_{H^2}\le \frac{4}{3}(a^\sharp)^{-1}(\lambda_k(A)+1)\|\psi\|_{L^2}\\ \textrm{for all}\ \psi\in S  }  }\ \  \max_{\varphi\in S\setminus\{0\}}\ \frac{\left\langle A\varphi,\varphi\right\rangle}{\|\varphi\|_{L^2}^2}
\label{min-max-A}   \end{equation}
   \end{proposition}
  \noindent{\it Proof of Proposition \ref{H2L2constraint}:}\ First note that the $\min\max$ in \eqref{min-max-A}, greater than or equal to $\lambda_k(A)$, given by \eqref{min-max}. We claim  $\lambda_k$ is  
  achieved at an eigenfunction, $\psi_k\ne0$, with 
  \[ \psi_k\in S_k(A)=\textrm{span of the first $k$ eigenfunctions of $A$, } \]
  where any $\psi\in S_k(A)$ satisfies:
  \begin{equation}
  \|\psi\|_{H^2}\le\frac{4}{3}(a^\sharp)^{-1}(\lambda_k(A)+1)\|\psi\|_{L^2}\ .
  \label{claimed-prop}
  \end{equation}
  Indeed, for any $\psi\in S_k(A)$, the span of the first $k-$ eigenfunctions of $A$, we have
  \[\|A\psi\|_{L^2}\le\lambda_k(A)\ \|\psi\|_{L^2}.\]
  It follows from \eqref{TTa} that
\[\frac{3}{4}a^\sharp\|\psi\|_{H^2}\le \|A\psi\|_{L^2}+\|\psi\|_{L^2}\le\left(\lambda_k(A)+1\right)\|\psi\|_{L^2}\ .\]
Thus, any $\psi\in S_k(A)$  satisfies \eqref{claimed-prop}.  Furthermore, the maximum of the quotient 
$\left\langle A\psi,\psi\right\rangle / \|\psi\|_{L^2}^2$ over $S_k(A)\setminus\{0\}$ is equal to $\lambda_k(A)$ and  is attained at $\psi_k$.  This completes the proof
  of Proposition \ref{H2L2constraint}. \medskip
  
  Continuing with the proof of Theorem \ref{lip-continuity}, 
 take $S$ to be any subspace of dimension $k$ and such that 
 \[v\in S\ \ \implies\ \ \|v\|_{H^2}\le \frac{4}{3}(a^\sharp)^{-1}(\lambda_k(A)+1)\|v\|_{L^2}.\] 
For all $0\ne v\in S$, we have
  \begin{align}
  \frac{\left\langle Tv,v\right\rangle_{L^2}}{\|v\|_{L^2}^2}\ &=\ 
   \frac{\left\langle \tT v,v\right\rangle_{L^2}}{\|v\|_{L^2}^2} + \frac{\left\langle (T-\tT)v,v\right\rangle_{L^2}}{\|v\|_{L^2}^2}\nn\\
   &\ge\ \frac{\left\langle \tT v,v\right\rangle_{L^2}}{\|v\|_{L^2}^2} - \frac{\|T-\tT\|_{H^2\to L^2}\|v\|_{H^2}\|v\|_{L^2}}{\|v\|_{L^2}^2}\nn\\
   &\ge\ \frac{\left\langle \tT v,v\right\rangle_{L^2}}{\|v\|_{L^2}^2} - \|T-\tT\|_{H^2\to L^2}\cdot
    \frac{\frac{4}{3}(a^\sharp)^{-1}(\lambda_k(T)+1)\|v\|_{L^2}^2}{\|v\|_{L^2}^2}\nn\\
    &=\ \frac{\left\langle \tT v,v\right\rangle_{L^2}}{\|v\|_{L^2}^2} - \|T-\tT\|_{H^2\to L^2}\cdot
    \frac{4}{3}(a^\sharp)^{-1}(\lambda_k(T)+1)
  \end{align}
  Therefore,
  \begin{align}
  \lambda_k(T)\ \ge\ \lambda_k(\tT)\ - \|T-\tT\|_{H^2\to L^2}\cdot
    \frac{4}{3}(a^\sharp)^{-1}(\lambda_k(T)+1)\ ,
    \label{estT-ge-tT}\end{align}
     thanks to \eqref{min-max-A} with $A=T$ and \eqref{min-max} with $A=\tT$.
  Interchanging $\tT$ and $T$ in  the above argument yields
  \begin{align}
  \lambda_k(\tT)\ \ge\ \lambda_k(T)\ - \|T-\tT\|_{H^2\to L^2}\cdot
    \frac{4}{3}(a^\sharp)^{-1}(\lambda_k(\tT)+1)\ .
    \label{est-tT-ge-T}\end{align}

\nit Estimates \eqref{estT-ge-tT} and \eqref{est-tT-ge-T} imply that
\begin{align}
 \left| \lambda_k(T)\ -\ \lambda_k(\tT) \right|\ &\le\ 
  \frac{4}{3}(a^\sharp)^{-1}(\lambda_k(T)\ +\ \lambda_k(\tT)+1)\cdot \|T-\tT\|_{H^2\to L^2}\nn\\
  &\le\ \frac{4}{3}(a^\sharp)^{-1}(2\lambda_k(T)\ +\ 1)\cdot \|T-\tT\|_{H^2\to L^2}\nn\\
  &\qquad\qquad\quad +\ 
  \frac{4}{3}(a^\sharp)^{-1}\cdot \|T-\tT\|_{H^2\to L^2}\ \left| \lambda_k(T)\ -\ \lambda_k(\tT) \right|
  \end{align}
  Therefore, since we have assumed $\|T-\tT\|_{H^2\to L^2}\le\frac{1}{4}a^\sharp$, we find that
  \begin{equation}
 \frac{2}{3}\ \left| \lambda_k(T)\ -\ \lambda_k(\tT) \right|\ \le\ \frac{4}{3}(a^\sharp)^{-1}(2\lambda_k(T)\ +\ 1)\cdot \|T-\tT\|_{H^2\to L^2}
 \nn\end{equation}
 This completes the proof of the Lipschitz bound \eqref{Lip-est-TmtT} and therewith Proposition \ref{lip-continuity}.\medskip
  
  Finally, Proposition \ref{ktomuk} is an immediate consequence of Proposition \ref{lip-continuity}.

\bibliographystyle{siam}
\bibliography{fw-dirac}

\begin{thebibliography}{10}

\bibitem{Ablowitz-Curtis-Zhu:12}
{\sc M.~Ablowitz, C.~Curtis, and Y.~Zhu}, {\em On tight-binding approximations
  in optical lattices}, Stud. Appl. Math, 129 (2012), pp.~362---388.

\bibitem{ANZ:09}
{\sc M.~Ablowitz, S.~Nixon, and Y.~Zhu}, {\em Conical diffraction in honeycomb
  lattices}, Physical Review A, 79 (2009), p.~053830.

\bibitem{Ablowitz-Zhu:11}
{\sc M.J. Ablowitz and Y.~Zhu}, {\em Nonlinear waves in shallow honeycomb
  lattices}, SIAM J. Appl. Math., 72 (2012).

\bibitem{Allaire-05}
{\sc G.~Allaire and A.~Piatnitski}, {\em Homogenization of the {\SH}~equation
  and effective mass theorems}, Comm. Math. Phys., 258 (2005), pp.~1--22.

\bibitem{avron-simon:78}
{\sc J.E. Avron and B.~Simon}, {\em Analytic properties of band functions},
  Annals of Physics, 110 (1978), pp.~85--101.

\bibitem{Segev-etal:08}
{\sc O.~Bahat-Treidel, O.~Peleg, and M.~Segev}, {\em Symmetry breaking in
  honeycomb photonic lattices}, Optics Letters, 33 (2008).

\bibitem{Berry-Jeffrey:07}
{\sc M.V. Berry and M.R. Jeffrey}, {\em Conical Diffraction: Hamilton's
  diabolical point at the heart of crystal optics}, Progress in Optics, 2007.

\bibitem{Courant-Hilbert-I}
{\sc R.~Courant and D.~Hilbert}, {\em Methods of {M}athematical {P}hysics},
  Interscience Publishers, Inc., New York, N.Y., 1953.

\bibitem{Eastham:73}
{\sc M.S. Eastham}, {\em The {S}pectral {T}heory of {P}eriodic {D}ifferential
  {E}quations}, Scottish Academic Press, Edinburgh, 1973.

\bibitem{FW:12}
{\sc C.L. Fefferman and M.I. Weinstein}, {\em Honeycomb lattice potentials and
  {D}irac points}, J. Amer. Math. Soc., 25 (2012), pp.~1169--1220.

\bibitem{goerbig:11}
{\sc M.O. Goerbig}, {\em Electronic properties of graphene in a strong magnetic
  field}.
\newblock arXiv:1004.3396v4.

\bibitem{Haldane:08}
{\sc F.D.M. Haldane and S.~Raghu}, {\em Possible realization of directional
  optical waveguides in photonic crystals with broken time-reversal symmetry},
  Phys. Rev. Lett., 100 (2008), p.~013904.

\bibitem{Indik-Newell:06}
{\sc R.~A. Indik and A.~C. Newell}, {\em Conical refraction and nonlinearity},
  Optics Express, 14 (2006), pp.~10614--10620.

\bibitem{Kuchment-01}
{\sc P.~Kuchment}, {\em The {M}athematics of {P}hotonic {C}rystals, in
  "{M}athematical {M}odeling in {O}ptical {S}cience"}, Frontiers in Applied
  Mathematics, 22 (2001).

\bibitem{RMP-Graphene:09}
{\sc A.H.~Castro Neto, F.~Guinea, N.M.R. Peres, K.S. Novoselov, and A.K. Geim},
  {\em The electronic properties of graphene}, Reviews of Modern Physics, 81
  (2009), pp.~109--162.

\bibitem{Segev-etal:07}
{\sc O.~Peleg, G.~Bartal, B.~Freedman, O.~Manela, M.~Segev, and D.N.
  Christodoulides}, {\em Conical diffraction and gap solitons in honeycomb
  photonic lattices}, Phys. Rev. Lett., 98 (2007), p.~103901.

\bibitem{Rechtsman-etal:12}
{\sc M.C. Rechtsman, J.M. Zeuner, Y.~Plotnik, Y.~Lumer, S.~Nolte, M.~Segev, and
  A.~Szameit}, {\em Photonic floquet topological insulators}.

\bibitem{RS4}
{\sc M.~Reed and B.~Simon}, {\em Modern {M}ethods of {M}athematical {P}hysics,
  {I}{V}}, Academic Press, 1978.

\bibitem{Wilcox:78}
{\sc C.~Wilcox}, {\em Theory of {B}loch {W}aves}, J. A'nalyse Math.,  (1978).

\end{thebibliography}
\end{document}